\documentclass[
  pre,
  reprint,
  nofootinbib,
  floatfix,
  aps,
]{revtex4-2}

\usepackage[english]{babel}
\usepackage{graphicx}
\usepackage{xcolor}

\usepackage{amsmath,amssymb}
\usepackage{multirow}

\usepackage{gensymb}      

\usepackage[normalem]{ulem} 

\usepackage{hyperref}
\usepackage{cleveref}

\begin{document}

\title{Bayesian Reasoning for  Physics Informed Neural Networks}

\author{Krzysztof M. Graczyk} 

\affiliation{Institute for Theoretical Physics, University of Wroc\l aw, plac Maxa Borna 9,
50-204, Wroc\l aw, Poland}

\author{Kornel Witkowski} 
\affiliation{Institute of Low Temperature and Structure Research\\
	Polish Academy of Sciences\\
ul.~Ok\'olna~2, Wroc\l aw, 50-422, Poland}

\begin{abstract}
We introduce an evidence-driven Bayesian formulation of physics-informed neural networks that enables automatic optimization of loss weights between PDE residuals, boundary conditions, and observational data. Unlike existing Bayesian PINN approaches based on sampling or variational inference, the proposed method uses a Laplace approximation to compute model evidence analytically, enabling efficient hyperparameter tuning and model comparison without posterior sampling. We demonstrate the method on the heat, wave, and Burgers’ equations, obtaining solutions in agreement with exact or reference results. In the Burgers’ equation example, we further show that the framework naturally integrates information from governing equations and noisy measurements, providing predictive uncertainties within a unified Bayesian setting.
\end{abstract}

	\maketitle

\section{Introduction}

The deep learning (DL) revolution impacts almost every branch of life and sciences~\cite{LeCun2015}. Deep learning models are also adapted and developed to solve problems in physics~\cite{MEHTA20191}. The statistical analysis of the data is one of the popular and straightforward domains of application of neural networks in physics~\cite{Graczyk:2010gw,Graczyk:2014lba}. Another one is using DL-like models to optimize and modernize computational systems in physics, starting with solvers for fluid flow \cite{Graczyk:2020hih,PhysRevFluids.3.074602} and ending with Monte Carlo generators in particle physics \cite{Otten:2019hhl,Bonilla:2025rrn,bonilla2025generativeadversarialneuralnetworks}. The generalization capabilities of deep neural networks have the potential to revolutionize physics by helping to fill gaps in our understanding of physical reality \cite{graczyk2024electronnucleuscrosssectionstransfer}.

Deep neural networks are examples of artificial neural networks (NNs) studied and utilized in various branches of life and science for years \cite{Sejnowski18}. In this paper, we consider feed-forward shallow NNs with one or at most two layers of hidden units used to solve partial differential equations (PDEs). 

Differential equations are the essence of physics. Indeed, they describe the system and its evolution in time in classical mechanics, electrodynamics, fluid physics, quantum mechanics, Etc. Some can be solved analytically, but a vast class of problems can be solved numerically only.

One of the exciting ideas is to adapt a neural network framework to solve PDEs numerically. The problem of numerical integration comes down to the optimization problem. Indeed, the approximate solution of a differential equation is parametrized by a feed-forward neural network that depends on the parameters  (weights). The optimal solution minimizes a cost function defined for a given equation. One of the first successful formulations of this idea was provided by Lagaris et al. \cite{Lagaris712178}. They applied the method to ordinary and partial differential equations. It was assumed that the feed-forward neural network, modified to satisfy initial or boundary conditions, was a solution of an equation. A similar idea was exploited by Lagaris \textit{et al.} for solving eigenvalue problems in quantum mechanics~\cite{LAGARIS19971}. 

One of the problems of the Lagaris \textit{et al.} approach was its relatively low universality, i.e., for each differential equation, one has to encode the initial or boundary conditions in the network parametrization. Although the final solution exactly satisfies the required requirements, the training of such a network could be smoother. Therefore, in modern applications, the boundary or initial conditions are introduced as additional contributions to the cost function. As a result, the boundary or initial constraints are satisfied by the obtained solution only on an approximate level. However, the optimization goes much smoother than in the Lagaris \textit{et al.} approach. This new approach was formulated by Raissi \textit{et al.} \cite{raissi2017physicsI,raissi2017physicsII,RAISSI2019686} and called \textit{Physics Informed Neural Network} (PINN). Two classes of problems were discussed:  data-driven solutions and data-driven discovery of PDEs. The PINN method has been popular, and it has been exploited for forward and inverse types of the problems \cite{Mishra_2021,mishra2021estimates,mishra2021estimates,Sirignano_2018,Lu_2021,Haghighat2020,thuerey2021pbdl,hao2023pinnacle,Jiang2022,Subramanian22,PDE-Net2017}. An extensive literature review on PINNs is given by Cuomo \textit{et al.} \cite{cuomo2022scientific}.

Karniadakis \textit{et al.} \cite{KarniadakisPINNreview} provides a comprehensive review of the PINN approach. In particular, they point out the major difficulties of the approach, namely, convergence to the global minimum, hyperparameter tuning, and fixing relative weights among the terms in the loss function. In particular, the last problem is addressed in this paper. The PINN idea can be naturally extended to a broader class of problems than PDEs, in which the NNs or machine learning systems are adapted to solve the problem or to optimize the system that solves the problem, see reviews by Faroughi \textit{et al.}~\cite{faroughi2023physicsguided} (the physics guided, informed, and encoded neural network approaches),  Meng \textit{et al.} \cite{Meng2022} and Hao \textit{et al.} \cite{hao2023physicsinformed} (physics informed machine learning including PINNs).

Artificial NNs have been studied for more than sixty years \cite{Rosenblatt62,hertz2018introduction,Goodfellow-et-al-2016,Sejnowski18}. Before the DL revolution, shallow feed-forward neural networks had been extensively studied. However, many of the achievements for shallow networks do not apply to deep neural networks because of the complexity of the DL systems. Nevertheless, even for the small NNs, there were several difficulties that one had to face, namely, overfitting problem, choice of the proper network architecture\footnote{The choice of the network architecture seems to be even more important for shallow than deep NNs.} and  fixing hyperparameters, and estimation of the uncertainties in network predictions. Bayesian statistics offer systematic methods that allow one to overcome these difficulties. Hence, in the nineties of the XX century, the Bayesian methods were discussed and adapted for artificial NNs~\cite{Bishop_book}.

Usually, in data analysis, the methods of frequentist rather than Bayesian statistics are utilized. There are essential differences between the two methodologies \cite{DAgostini_book}. The Bayesian approach seems more subjective in its formulation than the frequentist approach. However, the way the Bayesian framework  is defined allows a user to control the subjectivity of assumptions. From that perspective, Bayesian statistics can be used to study the impact of the model assumptions on the analysis results~\cite{Jeffreys_book}.

In Bayesian statistics, one must evaluate the posterior probabilities. The probability is a measure of the degree of belief that an event will occur \cite{DAgostini_book}. The initial (prior assumptions) are encoded in the prior probabilities. Bayes' rule is used to obtain the posterior probabilities. Note that the posterior probability can be continuously updated as new data or information arrives. Moreover, within the Bayesian approach, it is possible to consider various types of uncertainties \cite{survey_uncertainty_in_dnn}, namely, uncertainty due to data noise (aleatoric), uncertainty arising from lack of knowledge of the model parameters, and model dependence of the assumptions (epistemic).

Ultimately, it is possible to quantitatively and, in some sense, objectively compare various models and choose the one that is the most favorable to the data. Eventually, the Bayesian approach embodies Occam's razor principle \cite{Jeffreys_book,Gull1988}. The models that are too complex (with too many parameters, in the case of a neural network with too many units) are naturally penalized. Hence, the models obtained in the BF should not tend to overfit the data.

When the PDEs are solved with the PINN method, initial or boundary conditions can be naturally understood as prior assumptions. Moreover, the BF allows one to infer from the data the parameters that parameterize PDEs or, directly, the form of the PDEs. Eventually, the Bayesian PINN solution is provided with uncertainties. Therefore, the Bayesian approach to the PINN is one of the directions explored in the last few years \cite{PSAROS2023111902}.    

Indeed, there are many papers where various algorithmic formulations of the Bayesian PINNs are given \cite{Zhu_2018,Besginow2022}. The initial works concerned adapting the Gaussian process technique \cite{Rasmussen_Willimas_book} for regression \cite{RAISSI2017683,raissi2017inferring} -- the infinitely wide one-hidden layer network can be understood as a Gaussian process \cite{Neal:1995}. In many applications to obtain the posterior distributions, the Hamiltonian Monte Carlo and the variational inference are adapted \cite{Yang_2021,bonneville2021bayesian,more2023bayesian,NatureMachineIntelligenceDeepONet}. The broad class of Bayesian algorithms utilized for training NNs is discussed in a survey by Magris and Iosifidis~\cite{Magris_Bayesian_Survey_2022}. 

Note that the Bayesian PINN formulations relying on sampling-based posterior inference (e.g., HMC or variational methods) become computationally prohibitive even for moderately sized networks. At the same time, practical PINN performance is highly sensitive to the choice of loss-function weights, which are typically selected heuristically. This work addresses this gap by introducing an evidence-based BF that treats loss weights as inferable hyperparameters, enabling automated calibration without posterior sampling.

In the present paper, we consider the Bayesian framework (BF) for neural networks  \cite{MacKay_thesis,MacKay1992.4.3.415,MacKay1992.4.3.448,Bishop_book} whose works were motivated by results of Gull \cite{Gull1988,Gull1989} and Skilling \cite{Skilling1991}. 
In this approach,  all necessary probabilities are expressed in the Gaussian form. Indeed, the Laplace approximation is imposed to evaluate the posterior distribution. As a result, all probability densities are obtained in analytic form. The so-called evidence is estimated, a measure that classifies the models. The framework enables us to objectively select the optimal network architecture for a given problem, compare distinct network models, estimate uncertainties in network predictions and weights, and fine-tune the model's hyperparameters. In the past, we have adapted MacKay's framework to study the electroweak structure of the nucleons \cite{Alvarez-Ruso:2018rdx,Graczyk:2015kka,Graczyk:2014coa,Graczyk:2014lba,Graczyk:2011kh,Graczyk:2010gw}. For that project, we developed our own C++ Bayesian neural network library. In contrast, in the present project, we utilize the PyTorch environment \cite{NEURIPS2019_9015}, and both network weights and all hyperparameters are differentiable. It enables us to propose a significant improvement to the approach by making online hyperparameter updates more efficient.

We show that, by adapting BF for PINNs, one can numerically solve  heat, wave, and Burger's PDEs. We propose modifications to the original BF's algorithm. Indeed, while several Bayesian PINN formulations exist, most rely on sampling-based inference (e.g., Monte Carlo-like) or variational approximations, which can be computationally expensive. In contrast, the present work adapts the classical evidence framework of MacKay to PINNs, enabling analytic evaluation of model evidence and efficient hyperparameter optimization. This provides a principled mechanism for tuning relative loss weights and selecting model architectures without posterior sampling.
Note that the relative weights dictate the balance between various loss terms, including the PDE residuals, boundary and initial conditions, and the data contribution. As discussed by Wang et al. \cite{wang2020pinns} and others~\cite{KarniadakisPINNreview,faroughi2023physicsguided}, an improper weighting scheme can cause a single loss term to dominate the gradient, often leading to model failure. While the most straightforward remedy is manual tuning based on trial-and-error pre-runs~\cite{NVIDIA_2022}, significant research has shifted toward developing more consistent and automated solutions.

The paper is organized as follows: after the introduction in Sec. \ref{Sec:PINN}, we describe 
PINNs briefly, as well as discuss various approaches to fixing relative loss weights and overfitting problem. Our BF is formulated in Sec. \ref{Sec:BF}. In Sec. \ref{Sec:NumRes}, we show the comparison of HMC and this paper Bayesian framework to nonlinear regression problem (Sec. \ref{Sec:HMC}), as well as in Sections \ref{Sec:NumRes:Heat}, \ref{Sec:NumRes:wave} and \ref{Sec:NumRes:Burgers}, we present the solutions of heat, wave, and Burger's equations, respectively. In Sec.~\ref{Sec:Summary}, we summarize. The paper contains two appendices discussing the details of the Bayesian approach.

\section{PINN}

\label{Sec:PINN}

\subsection{Multilayer perceptron configuration}

\label{Sec:PINN:NeuralNetworks}

\begin{figure}
	\centering
	\includegraphics[width=0.5\textwidth]{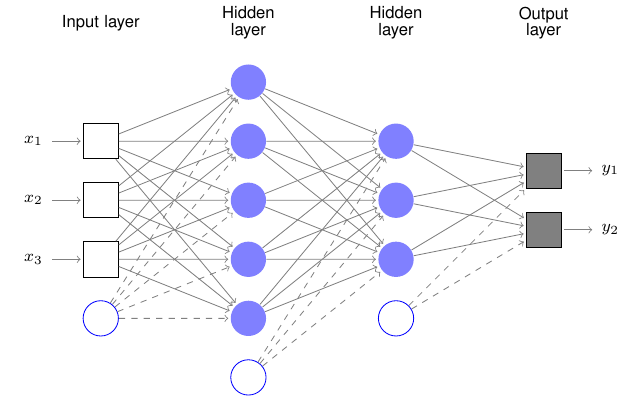}
	\caption{The above MLP contains two hidden unit layers. Empty squares denote the input, whereas filled circles represent the output; Blue filled and open circles indicate hidden and bias units, respectively.  \label{Fig_2MLP_graph}}    
\end{figure}
We shall consider the feed-forward neural network in the multilayer perceptron (MLP) configuration. The MLP, denoted by $\mathcal{N}$,  represents a nonlinear map from the input space of the dimension~$d_{in}$ to the output space of the dimension~$d_{out}$,
\begin{equation}
\mathcal{N}: \mathbb{R}^{d_{in}} \mapsto \mathbb{R}^{d_{out}}.
\end{equation}
Note that $\mathcal{N} = \mathcal{N}(\mathbf{x},\textbf{w})$ is a function of input vector $\mathbf{x} \in \mathbb{R}^{d_{in}}$ and its parameters called weights, denoted by the vector $\textbf{w} = (w_1,\dots, w_i,\dots, w_W)$, where $W$ is the total number of weights in a given network.

The MLP is represented by a graph consisting of several layers of connected units (nodes). Each edge of the graph corresponds to a function parameter, a weight, $w_i$. A unit (node) in a given layer is a single-valued function (activation function). Note that every unit, except the bias unit, is connected with the nodes from the previous layer. In a graph representing a neural network, we distinguish an input (input units), hidden layers of units, and an output layer. Each layer (except the output layer) has a bias unit connected only with the following layer units.   It is a single-valued constant function equal to one. An example of a graph of a network with three-dimensional input and two-dimensional output is given in Fig.~\ref{Fig_2MLP_graph}. This network has two hidden layers with five and three hidden units. The properties of the network, $\mathcal{N}$, are determined by its structure (architecture -- topology of the connections) and the type of activation functions that settle in the neurons (units).

The universal approximation theorem~\cite{Cybenko_Theorem,Hornik89,FUNAHASHI1989183} states that a feed-forward neural network with at least one layer of hidden units can approximate well any given continuous function. It is a fundamental property of NNs, widely used in nonlinear regression and classification problems.

\subsection{PINN:  a formulation}

\label{Sec:PINN:PINN}

We formulate the PINN approach similarly as in Ref.~\cite{NVIDIA_2022}. Let us consider the sets of equations:
\begin{eqnarray}
\label{Eq:Differential_equation}
{Eq}_i(\mathbf{x}, \hat{f}, \partial \hat{f},\partial^2 \hat{f} ) & = & 0, \quad i=1,2, \dots, q, \;\mathbf{x} \in \mathcal{M}, \\
\label{Eq:Differential_equation_boundary}
\mathcal{B}_j(\mathbf{x}, \hat f , \partial \hat f) &=& 0, \quad j=1,2, \dots, b, \; \mathbf{x} \in \partial \mathcal{M},
\end{eqnarray}
where ${Eq}_i$'s and $\mathcal{B}_j$'s are the differential and boundary operators respectively; $\mathcal{M} \subseteq \mathbb{R}^d$, $d=1,2,\dots$;  
$\hat f$~is the solution of the equations (\ref{Eq:Differential_equation})-(\ref{Eq:Differential_equation_boundary}). Note that $\partial \hat{f}$ and $\partial \partial \hat f$ refer to any first-order partial derivative and second-order derivatives with respect to the input variables, respectively.

In the simplest version of the PINN approach, it is assumed that an approximate solution of the (\ref{Eq:Differential_equation}) with 
boundary conditions (\ref{Eq:Differential_equation_boundary}) is given by neural network $\mathcal{N}(\mathbf{x};\mathbf{w})$.  Then, the network solution, $\mathcal{N}(\mathbf{x}; \mathbf{w}^*)$ is so that 
\begin{equation}
\mathbf{w}^* = \arg \min_{\mathbf{w}} E , 
\end{equation}
where 
\begin{widetext}
\begin{equation}
E   =  \frac{1}{2}\left\{\sum_{i=1}^{q}  \beta_q^{i}\sum_{k=1}^{N_q} \left[{Eq}_i(\mathbf{x}_k;\mathcal{N},\partial \mathcal{N}, \partial^2 \mathcal{N})\right]^2 
+ 
\sum_{j=1}^{b}\beta^{j}_b \sum_{k=1}^{N_b} \left[\mathcal{B}_i(\mathbf{x}_k;\mathcal{N},\partial \mathcal{N}, \partial^2 \mathcal{N})\right]^2\right\},
\label{Eq:loss_first_time}
\end{equation}
where $\partial \mathcal{}$ and $\partial^2 \mathcal{}$ refer to any derivative with respect to input varaibales $\mathbf{x}$. 
\end{widetext}
The parameters $\beta^{i}_q$'s and $\beta^{j}_b$'s control the relative weights of the components of the loss $E$. To simplify the discussion, we shall consider only one $\beta^{i}_q = \beta_q$ and 
one $\beta^{j}_b= \beta_b$. We keep them together in the vector  $\boldsymbol \beta = (\beta_q,\beta_b)$. Moreover, in this paper, we consider the PDEs with $q=1$. Hence, the loss consists of 
\begin{equation}
\label{Eq:Edq}
E_D^{q}  = \frac{1}{2}\sum_{k=1}^{N_{q}} \left( Eq(\mathbf{x}_k;\mathcal{N},\partial \mathcal{N}, \partial^2 \mathcal{N})\right)^2 
\end{equation}
and
\begin{equation}
\label{Eq:Edb}
E_D^{b}  = 
\frac{1}{2}\sum_{j=1}^b \sum_{k=1}^{N_{b}} \left( 
\mathcal{B}_j(\mathbf{x}_k; \mathcal{N},\partial \mathcal{N}, \partial^2 \mathcal{N}) 
\right)^2,
\end{equation}
where
\begin{equation}
E  =  \beta_q E_D^{q}+\beta_b E_D^{b}
\end{equation}

By $N_{q/b}$, we denote the number of points in $\mathcal{M}/\partial \mathcal{M}$ collected to optimize the network model.  The total training dataset consists of points from $\mathcal{M}$ and $\partial \mathcal M$, namely,  $\mathcal{D} = \mathcal{D}_q \cup \mathcal{D}_b $. Eventually, we introduce the vector $\boldsymbol \beta = (\beta_q, \beta_b,... )$.

\subsection{Overfitting}

The increasing number of units in the NN broadens its adaptive abilities. However, if a network is large (contains many hidden units), it may overfit the data and, as a result, have poor generalization. On the other hand, if the network is small (few hidden units), it tends to underfit the data. This dilemma is known as the bias-variance trade-off~\cite{Geman.1992.4.1.1}. There are some rules of thumb for solving the problem of under- and over-fitting. The most popular is to constrain the values of the weight parameters by adding to the loss (\ref{Eq:loss_first_time}) the penalty term of the form:
\begin{equation}
\label{Eq:penalty_w_ini}
E_w = \frac{\alpha}{2} \sum_{i=1}^W w_i^2,
\end{equation}
where $\alpha$ is the decay width parameter.

There are also other popular methods for dealing with overfitting. However, we shall utilize the one described above because of its Bayesian interpretation. We shall consider several decay width parameters accommodated in the vector $\boldsymbol \alpha$. Indeed, we consider the penalty term of the form:
\begin{eqnarray}
E_w &=& \sum_{i=1}^C \alpha_i E_w^i \\
E_w^i &=& \frac{1}{2} \displaystyle\sum_{k = s_{i-1}}^{ s_{i}} w_{k}^2, 
\quad s_i =  \sum_{l=1}^{i} W_l, \; s_0=1
,
\end{eqnarray}
where, $C$ denotes the number of weight classes; $w_j$ refers to the $j$-th weight, $W_i$ is the number of weights in $i$-th class fo weights; $\alpha_i$ is a decay width for $i$-th class of weights. The vector $\boldsymbol \alpha = (\alpha_1,\dots,\alpha_C)$ contains all decay weights.   The weights in a given layer form a separate class, similar to the bias weights in each layer.

\section{Bayesian framework for neural networks}
\label{Sec:BF}

Solving the PDE in the PINN approach comes down to optimization tasks. Namely, it is assumed that the neural network parameterizes the solution, then the loss, such as in Eq.~\ref{Eq:loss_first_time},  is formulated, and model parameters are optimized. Solving the PDE can be understood as the Bayesian inference process. Indeed, we assume prior knowledge about the solution and its parameters and then search for the posterior realization. Another problem is the proper fine-tuning of the beta parameters, which control the impact of the boundary conditions on the solution. The choice of optimal network architecture seems to be essential, too. Ultimately, it is evident that the neural network's solution is only an approximation of the \textit{true} solution, or can be treated as a statistical prediction of the desired solution. Therefore, the PINN solution should be provided with uncertainties. We shall argue in the next section that the Bayesian statistical methods can help to establish the optimal structure of the network architecture; moreover, with the help of Bayesian reasoning, one may obtain the optimal values for the hyperparameters such as $\beta_q$ and $\beta_b$, and $\alpha$  and compute uncertainties for the obtained fits.

\subsection{Implementation of Bayesian framework}

The main idea of our approach is to assume that all the densities should have a Gaussian-like shape. Then the Laplace approximation and Hessian-based inference~\cite{MacKay_thesis,MacKay1992.4.3.415,MacKay1992.4.3.448,Bishop_book} is performed. While this restricts applicability to shallow or moderately sized PINNs, it enables analytic evidence evaluation and efficient hyperparameter optimization. Extension to deep architectures would require alternative posterior approximations and is beyond the scope of this work.

Estimating optimal weights, hyperparameters, and model structure  can be split into three steps,  discussed in Appendix~\ref{Appendix_0}. Below, we summarize the main points of the approach. The detailed description of the approach is given in Appendix~\ref{Appendix_A}. 

\subsubsection{Step one}
From the first step of inference, one obtains the loss, which is used to obtainthe  maximum posterior (MP) configuration of weights, namely,
\begin{eqnarray}
\mathbf{w}_{MP} &=& \arg \min_{\mathbf{w}} E_T , \\
\label{Eq:ET_definition}
E_T &=& \sum_{i=1}^C \alpha_i E_w^i + \beta_{q} E_D^{q} + \beta_{b} E_D^{b}.
\end{eqnarray}

Before we review the next steps of the approximation, let us comment on the validity of the Gaussian approximation. In neural network analyses, the error function has many minima, some related to the symmetry property embodied by a particular type of network architecture. There are local minima that are not related by symmetry. As proposed in  \cite{MacKay1992.4.3.448,nasa_1991,Bishop_book}, one may consider every local minimum (maximum of the posterior), not related by symmetry, as a separate model configuration with its own optimal hyperparameter configuration. It may happen that distinct $\mathbf{w}_{MP}$ (not related by symmetry property) can have different hyperparameters $\boldsymbol{\alpha}_{MP}$, $\boldsymbol{\beta}_{MP}$. From that perspective, the integrals such as Eq.~\ref{Eq:posterior_alpha_beta} should be interpreted as an integral over the neighborhood of the optimal parameters of the local minimum. In such an approximation, one collects parameter and hyperparameter configurations corresponding to some local maxima of posterior densities.

\subsubsection{Step two}
From the second step approximation, by applying the so-called evidence approximation~\cite{Gull1988,MacKay1992.4.3.415}, one obtains
\begin{widetext}
\begin{eqnarray}
\ln  p(\mathcal{D}\mid \boldsymbol\alpha,\boldsymbol\beta,\mathcal{N}) &\approx & - \sum_{i=1}^C \alpha_i E_w^i(\mathbf{w}_{MP}) 
- \beta_{q} E_D^{q}(\mathbf{w}_{MP}) - \beta_{b} E_D^{b}(\mathbf{w}_{MP})
- \frac{1}{2}\ln |\mathbf{A}| - \sum_{i=1}^C \frac{W_i}{2}\ln\left( \frac{2\pi}{\alpha_i}\right)  \nonumber \\
&  & \!\!\!\!\!\!\!\! -  \frac{N_{q}}{2} \ln \left( \frac{2\pi}{\beta_{q}} \right)
- \frac{N_{b}}{2} \ln \left( \frac{2\pi}{\beta_{b}}\right), 
\end{eqnarray}
\end{widetext}
where $\mathbf{A}$ is defined by
\begin{equation}
\mathbf{A}  = \mathbf{H} + \mathbf{I}_{W}({\boldsymbol\alpha}),
\end{equation}
where
\begin{equation}
\mathbf{I}_{W}({\boldsymbol\alpha}) = \mathrm{diag}(\underbrace{\alpha_1,\dots,\alpha_1}_{W_1\, times},\dots, \underbrace{\alpha_C,\dots,\alpha_C}_{W_C\, times})
\end{equation}
and 
\begin{equation}
\mathbf{H}_{ij} = \beta_{q} \nabla_i \nabla_j E_D^{q} + \beta_{b} \nabla_i \nabla_j E_D^{b}
\end{equation}
is the Hessian, $\nabla_i = \partial_{w_i}$.

One searches for the maximum of $p(\mathcal{D}\mid \boldsymbol \alpha,\boldsymbol \beta,\mathcal{N})$. In the original MacKay's framework, the hyperparameters are iterated according to the necessary condition for the maximum, namely, from the equations:
\begin{eqnarray}
\frac{\partial}{ \partial \boldsymbol\alpha} p(\mathcal{D}\mid \boldsymbol \alpha, \boldsymbol\beta,\mathcal{N}) &=& 0, \\
\frac{\partial}{ \partial \boldsymbol\beta} p(\mathcal{D}\mid \boldsymbol \alpha, \boldsymbol\beta,\mathcal{N}) &=& 0.
\end{eqnarray}
However,  the solution of the above equations is only approximate, and in the case of the PINN application, unstable during training.

Instead of this algorithm, we propose to consider an additional loss given by 
\begin{equation}
E_{hyp} = -\ln  p(\mathcal{D}\mid \boldsymbol \alpha,\boldsymbol \beta,\mathcal{N})
\end{equation}
and optimize it with respect to hyperparameters: $\boldsymbol \alpha$, $\boldsymbol \beta$. 

The loss, after neglecting constant terms, has the form

\begin{eqnarray}
E_{hyp}  &=&   \sum_{i=1}^C \alpha_i E_w^i(\mathbf{w}_{MP}) \nonumber \\
& & + \beta_{q} E_D^{q}(\mathbf{w}_{MP}) + \beta_{b} E_D^{b}(\mathbf{w}_{MP})
+ \frac{1}{2}\ln |\mathbf{A}|  
\nonumber \\
&     & 
-\sum_{i=1}^C \frac{W_i}{2}\ln\alpha_i
-  \frac{N_{q}}{2} \ln \beta_{q}
- \frac{N_{b}}{2} \ln \beta_{b} .
\label{Eq:HyperLoss_II}
\end{eqnarray}

The first type of parameter, $\boldsymbol \alpha$ controls the prior's impact on the posterior, and the optimal solution is the one that generalizes well. The $\mathcal \beta$ provide information about the noise in the data; however, this information is incomplete as long as the number of data points is small. On the other hand, the algorithm estimates the $\beta$s  to maximize the model's evidence. Hence, the algorithm balances various posterior contributions to obtain a model with the best generalization abilities (in this case, mostly interpolation).

\subsubsection{Step three}

In the third step of inference, one computes the evidence. It is a conditional probability $p(\mathcal{D}\mid \mathcal{N})$.  After the approximations described in the Appendix~\ref{Appendix_A},  the log of evidence reads
\begin{eqnarray}
\ln p(\mathcal{D}\mid \mathcal{N}) 
&=&     
\ln  p(\mathcal{D}\mid \boldsymbol\alpha,\boldsymbol\beta) + \ln \sigma_{\ln\alpha} \nonumber \\
& & +  \ln \sigma_{\ln\beta_{q}} + \ln \sigma_{\ln\beta_b}  + comb, 
\label{Eq:Evidence_general_Ia}
\end{eqnarray}
where
\begin{eqnarray}
\ln  p(\mathcal{D}\mid \boldsymbol\alpha,\boldsymbol\beta) &=& 
- \sum_{i=1}^C \alpha_i E_w^i(\mathbf{w}_{MP}) \nonumber \\
& & - \beta_{q} E_D^{q}(\mathbf{w}_{MP}) - \beta_{b} E_D^{b}(\mathbf{w}_{MP})
\nonumber \\
&     & - \frac{1}{2}\ln |\mathbf{A}|   +\sum_{i=1}^C \frac{W_i}{2}\ln\alpha_i
\nonumber \\
& & 
\label{Eq:final_}
+  \frac{N_{q}}{2} \ln \beta_{q}
+ \frac{N_{b}}{2} \ln \beta_{b} 
\end{eqnarray}
and
\begin{eqnarray}
\ln {\sigma_{\ln\alpha_i}} &=& -\frac{1}{2} \ln \left[ -
\alpha^2_i \frac{\partial^2 }{\partial \alpha_i^2} \ln p(\mathcal{D}\mid \boldsymbol\alpha, \boldsymbol\beta ) \right],
\\
\ln {\sigma_{\ln\beta_i}} &=& -\frac{1}{2} \ln \left[ -
\beta^2_i \frac{\partial^2 }{\partial \beta_i^2} \ln p(\mathcal{D}\mid \boldsymbol\alpha, \boldsymbol\beta ) \right].
\end{eqnarray}
The expression (\ref{Eq:Evidence_general_Ia}) consists of the symmetry contribution, as discussed in Ref. \cite{Chen1993}, namely, 
\begin{equation}
comb =  \sum_{i=1}^L \left( \ln M_i!  + M_i \ln 2\right),
\end{equation}
where $M_i$ denotes number of units in the $i$-th hidden layer, $L$ denotes number of hidden layers.

\subsection{Uncertainty for the network predictions}

The Bayesian approach to neural networks can take into account two types of uncertainties, namely, aleatoric and epistemic \cite{survey_uncertainty_in_dnn}. The first has roots in the data, the other in the model and its parameter dependence. In our case, we would rather concentrate on the epistemic uncertainties.  The Bayesian approach used here allows us to quantify two types of uncertainty: model uncertainty, described by the Bayesian evidence, and parameter uncertainty arising from the neural network weights. The latter can be computed in the covariance matrix approximation. Indeed,  the $1\sigma$ uncertainty due to the variation of the weights reads~\cite{Thodberg1993AceOB}
\begin{widetext}
\begin{equation}
\Delta^2 \mathcal{N}(\mathbf{x}_{q(b)},\mathbf{w}_{MP}) =   
\sum_{i,j}\nabla_i \mathcal{N}(\mathbf{x}_{q(b)},\mathbf{w}_{MP}) \; (\mathbf{A}^{-1})_{ij} 
\nabla_j \mathcal{N}(\mathbf{x}_{q(b)},\mathbf{w}_{MP}),
\end{equation}
\end{widetext}
where $\mathbf{x}_{q(b)}$ refers to point from $\mathcal M (\partial \mathcal M)$; $\nabla_i \equiv  \partial/\partial w_i$.

\subsection{Training algorithm}

Here, we briefly summarize the training algorithm.

\begin{itemize}
	
	\item Set the network architecture and initial values for the weights.
	
	\item Set initial values of 
	$\mathbf{\alpha}_{i}=\mathbf{\alpha}_i^{(0)}$ and $\mathbf{\beta}_a=\mathbf{\beta}_a^{(0)}$. 
	
	\item To get $\mathbf{w}^*$ minimize the loss (\ref{Eq:ET_definition}). 
	
	\item To get hyperparameters $\mathbf \alpha $ and $\mathbf \beta$ minimize the loss\footnote{We compute $\ln |\mathbf{A}|$ using PyTorch routines.} (\ref{Eq:HyperLoss_II}).
	
	\item The best network has the highest evidence value computed from (\ref{Eq:Evidence_general_Ia}).

	\item[*] In both optimizations Adam algorithm~\cite{kingma2017adam} was utilized. 
 
\end{itemize}

Note that during training, one must update both the weights and the hyperparameter values simultaneously. However, searching for the optimal hyperparameters' configuration for given $\mathbf{w}_{MP}$ makes numerical sense when one is close to the peak of the posterior. The Adam algorithm we use is rather slow in approaching the minimum. Therefore, we start weight optimization after a certain number of epochs. Eventually, we optimize the hyperparameters every several epochs; otherwise, the procedure can be unstable. In that case, we updated hyperparameters for every epoch. Note that the simultaneous weight and hyperparameter updates can be understood as a bi-level optimization problem~\cite{zhang2023introduction}.

\subsection{Relation to adaptive loss weighting in PINNs}
\label{Sec:Adaptive_losses_review}

As mentioned in the Introduction, fixing relative weights between losses is one of the problems of the PINN approach. This paper proposes an alternative approach grounded in Bayesian reasoning. While the Bayesian ingredients themselves are classical, their integration into a unified evidence-driven framework for tuning PINN loss components constitutes the main methodological contribution of this work.
\begin{table*}
\centering
\caption{Examples of PINN loss-balancing strategies. \label{Table_balancing}}
\begin{tabular}{|p{3.2cm}|p{4.5cm}|p{5cm}|}
\hline
\textbf{Category} & \textbf{Representative Work} & \textbf{Key Strategy} \\
\hline
Gradient-balancing / Heuristic 
& ReLoBRaLo; Dual-Balanced PINN; Multi-Objective Loss Balancing 
& Adaptive weighting based on gradient statistics or heuristic rules during training \\
\hline
Uncertainty / Likelihood-Based 
& ASW-PINN variants; Empirical NTK-based uncertainty methods 
& Weight inference via noise or uncertainty estimation with probabilistic interpretation \\
\hline
NTK-Informed / Conditioning 
& NTK-based self-adaptive PIPNNs; Adaptive lifting PINN 
& Loss scaling guided by NTK analysis or conditioning of training dynamics \\
\hline
\end{tabular}
\end{table*}

Let us distinguish three families of loss-weight balancing, namely, choosing the relative weights of the contributing losses:
\begin{enumerate}
\item[(i)] gradient-balancing heuristics (multi-task style) \cite{pmlr-v80-chen18a,BISCHOF2025117914,zhou2025dualbalancingphysicsinformedneuralnetworks, farea2025multiobjectivelossbalancingphysicsinformed}; 
\item[(ii)] uncertainty/likelihood-based weighting \cite{Xiang_2022,zhu2025twostageadaptiveliftingpinn}; 
\item[(iii)] Neural Tangent Kernel (NTK) informed weighting (conditioning the training dynamics) \cite{ZHANG2024117042}. 
\end{enumerate}
In Table~\ref{Table_balancing}, we summarize the discussed strategies below.

In the first family, (i), the idea is to choose weights so that gradients from each loss term are comparable, preventing one term from dominating. The first applications come from multitasking problems. One example is GradNorm~\cite{pmlr-v80-chen18a}, which dynamically adjusts task weights to equalize training rates via gradient norms. The PINN-specific variants include ''learning-rate annealing/gradient statistics'' style strategies motivated by gradient pathologies in PINNs~\cite{doi:10.1137/20M1318043}. It differs from our approach, as we propose  evidence-driven hyperparameter selection. Unlike our Bayesian evidence-based approaches, where relative loss weights are treated as inferable hyperparameters with a clear statistical interpretation, adaptive PINNs from category (i), such as Wight and Zhao \cite{wight2020solving}, rely on heuristic redistribution of training points to balance optimization difficulty, without providing uncertainty quantification or model-selection criteria. Another example of the first group approach is CoPhy-PGNN \cite{elhamod2021cophypgnnlearningphysicsguidedneural}, which represents a heuristic adaptive loss-balancing approach based on training-stage scheduling of competing loss terms, without statistical interpretation or evidence-based selection of relative weights. The method of Kim \textit{et al.}~~\cite{kim2020dpmnoveltrainingmethod} falls within heuristic adaptive PINN training strategies, where competing loss terms are balanced by explicitly controlling the physics residual during optimization rather than through probabilistic inference of loss weights or kernel-based conditioning.

A representative example of the second category (ii) is a self-adaptive lbPINNs (loss-balanced Physics-Informed Neural Networks), which treats each loss term as a Gaussian likelihood and updates weights via maximum likelihood estimation (MLE) during training \cite{Xiang_2022}. A key feature of this method is that the relative weights acquire a statistical interpretation—specifically, as inverse noise variances—rather than merely serving as importance factors. While our approach shares this statistical motivation, we utilize a Bayesian evidence framework based on maximum posterior approximation, providing a unified interpretation for all hyperparameters.

In the third category (iii), NTK-informed weighting (conditioning the training dynamics), the idea is to use NTK insights to choose weights that improve conditioning or equalize effective learning dynamics across terms.
One example would be NTK-based self-adaptive loss weighting inside a domain-decomposition / parallel PINN framework. \cite{ZHANG2024117042}. Another is the NTK perspective of Wang, Yu, and Perdikaris \cite{wang2020pinns},  which also  belongs to the NTK-informed conditioning methods in the PINN landscape, where pathology is diagnosed via neural tangent kernel analysis and loss-term scaling is guided by the NTK’s spectral properties to mitigate convergence-rate disparities among loss components.
Our approach is not NTK-based; it uses a Hessian-based object instead. Unlike gradient-balancing or NTK-based strategies, the present approach treats loss weights as Bayesian hyperparameters, selecting their values by maximizing the model evidence, thereby incorporating Occam’s razor and enabling model comparison.

\section{Numerical experiments}

\label{Sec:NumRes}

We will solve three types of partial differential equations: heat, wave, and Burger's. Analytic solutions are available for the first two equations, but there is no analytic solution for Burger's equation. To approach this problem, we have used the PyTorch framework \cite{NEURIPS2019_9015} and collected $100$ solutions for each of the three PDEs. We will be selecting the best solution based on the evidence. 
\begin{figure*}
	\centering
	\includegraphics[width=0.85\textwidth]{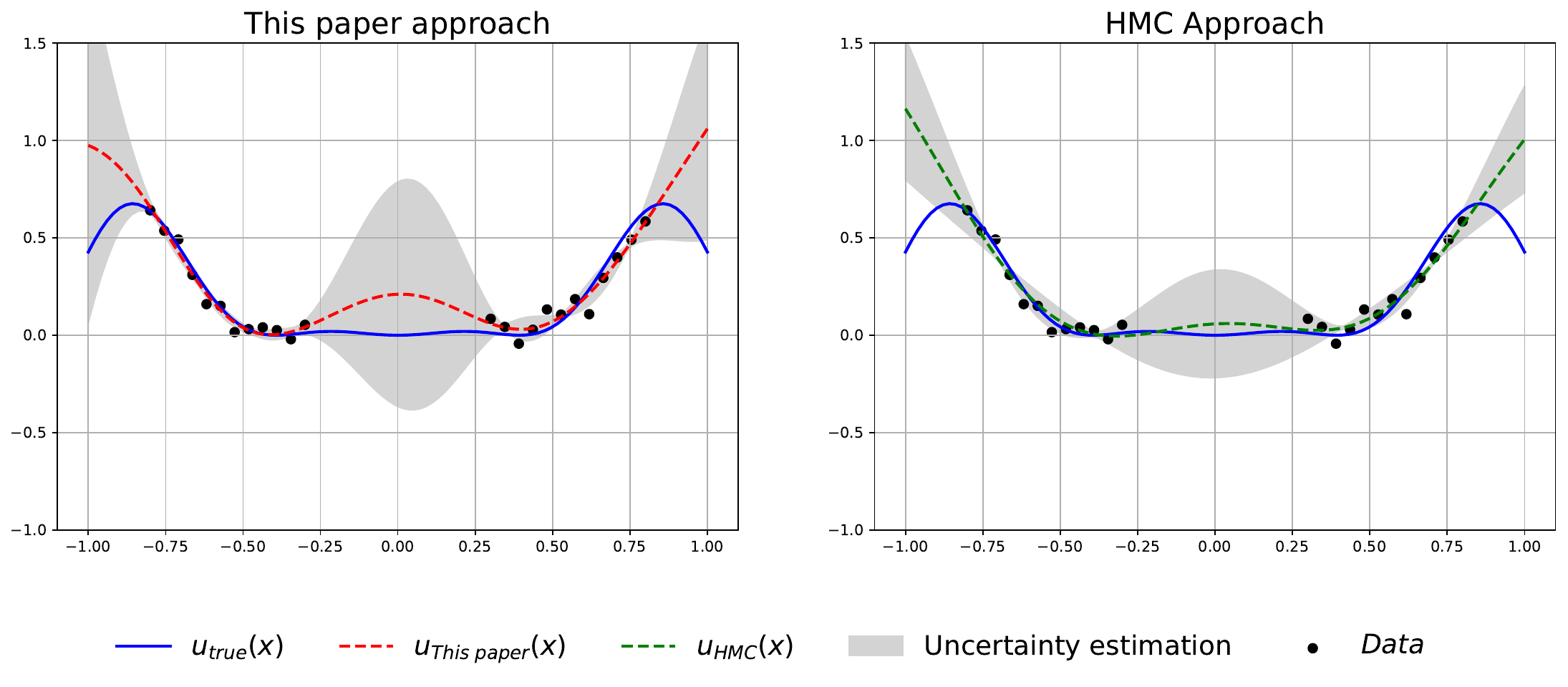}
	\caption{
    Nonlinear regression within this paper approach (left panel) and HMC approach (right panel). The red/blue line corresponds to this paper/HMC fit. The solid blue line describes the true value.
    \label{fig:BayesVsHMC}
	}
\end{figure*}

Before discussing the results for PDEs, we will begin the presentation by comparing our framework to the Bayesian approach based on the Markov chain algorithm. We will apply both methods to fit a curve. 
Next, we will simplify our approach to solving the heat equation by considering only two hyperparameters, $\alpha$ and $\beta$. We will repeat this scenario for the wave equation as well. Then, we solve the wave equation within the full model, keeping all allowed weight-decay parameters and beta values. Finally, we shall solve Burger's equation using the semi-simplified version of our approach, where we consider one weight-decay hyperparameter $\alpha$ and several beta parameters that control the equation, boundary conditions, and additional data contributions. Indeed, in the last example, we study a problem in which we also have additional information from the ''measured'' data. Namely, we add information about the solution, with some uncertainty, to the analysis. This additional contribution is provided with a corresponding $\beta$ hyperparameter, which our algorithm fixes.

\subsection{Comparison with Hybrid Monte Carlo}

\label{Sec:HMC}

Let us compare the results of the Bayesian approach introduced in the previous section with the Bayesian approach in which the posterior distributions are sampled using the Hybrid Monte Carlo (HMC) algorithm \cite{Neal:1995,Brooks_2011}. This approach is adapted in Ref. \cite{Yang_2021} to formulate the B-PINN framework. In the HMC approach, there is no need to make simplifications in evaluating the posteriors. However, as in the Bayesian framework presented in this paper, fixing the hyperparameters like $\alpha$ and $\beta$ requires additional care~\cite{Neal:1995}. In fact, these parameters are usually obtained from prior trial analysis by appropriate sampling or from grid analysis. We show in the following sections that our approach avoids sampling entirely and is therefore computationally advantageous for hyperparameter tuning.

To compare the predictions of both approaches, we consider the nonlinear regression problem. Let us consider the curve parametrized as follows
\begin{equation}
\label{Eq:Wave_b2_hmc}
u(x) = x^2\cos^2(4x).
\end{equation}
To generate the data we compute $u_{data}(x_i) = u(x_i) + \epsilon_i $, where $\epsilon_i$ is drawn from the normal distribution with zero mean and $0.05$ variance, whereas $x_i$ are obtained in the following way: $12$ points are drawn from the interval $[-0.8, -0.3]$ and $12$ points are drawn from the interval $[0.3, 0.8]$.

We fit the data with a one-hidden-layer neural network consisting of six hidden units. The hyperbolic tangent gives the activation functions. We trained a number of networks, each of them for $15000$ epochs. We show the best fit according to evidence. We switched on the Bayesian algorithm to evaluate hyperparameters after the $1000$th epoch and converged at $\alpha=0.24$, $\beta=596$. 

For the HMC approach, we take the same network architecture and hyperparameters $\alpha$ and  $\beta$ as in the previous analysis. To obtain the HMC fit, we drew from the posterior $15 000$ samples. The mean over collected samples (with $1000$ burn-in-steps) gives the model response as a final result. The corresponding variance determines the uncertainty.

Both fits are depicted in Fig.~\ref{fig:BayesVsHMC}. As we see, both approaches lead to similar fits and comparable uncertainties. However, one can expect that for more complex problems, the uncertainties computed from the HMC approach would be more prominent than in this paper's approach. Indeed, in our approach, we focus on the
local posterior peak, and assume that it has a Gaussian-like shape. 
On the other hand, in the HMC approach, there is no clear guideline for tuning hyperparameters. Additionally, the HMC algorithm's parameters, namely, the number of samples, and the hyperparameters, such as the number of leapfrog steps, require estimation.

\subsection{Heat equation}

\label{Sec:NumRes:Heat}

Let us consider the heat equation of the form
\begin{eqnarray}
\frac{\partial u}{\partial t} - \kappa \frac{\partial^2 u}{\partial x^2} & = & 0, 
\quad x \in [0,1], \;\; t \in [0, 1],
\label{Eq:Heat}
\\
u(0, t) & = & u(1, t)  =  0,
\label{Eq:Heat_b1}
\\
u(x, 0) & = &  \sin(\pi x)
\label{Eq:Heat_b2}
\end{eqnarray}
For the above case, the analytic solution reads  
\begin{equation}
u_{\text{true}}(x, t)= \sin(\pi x) \exp(-\kappa \pi^2 t).
\end{equation}

We consider a small network with one hidden layer to solve the heat equation (\ref{Eq:Heat}) with boundary conditions (\ref{Eq:Heat_b1}-\ref{Eq:Heat_b2}). The hidden layer has six neurons with hyperbolic tangent activation functions. In the output layer, there is no activation function.

\begin{figure}[h]
	\begin{center}
	\includegraphics[width=0.4\textwidth]{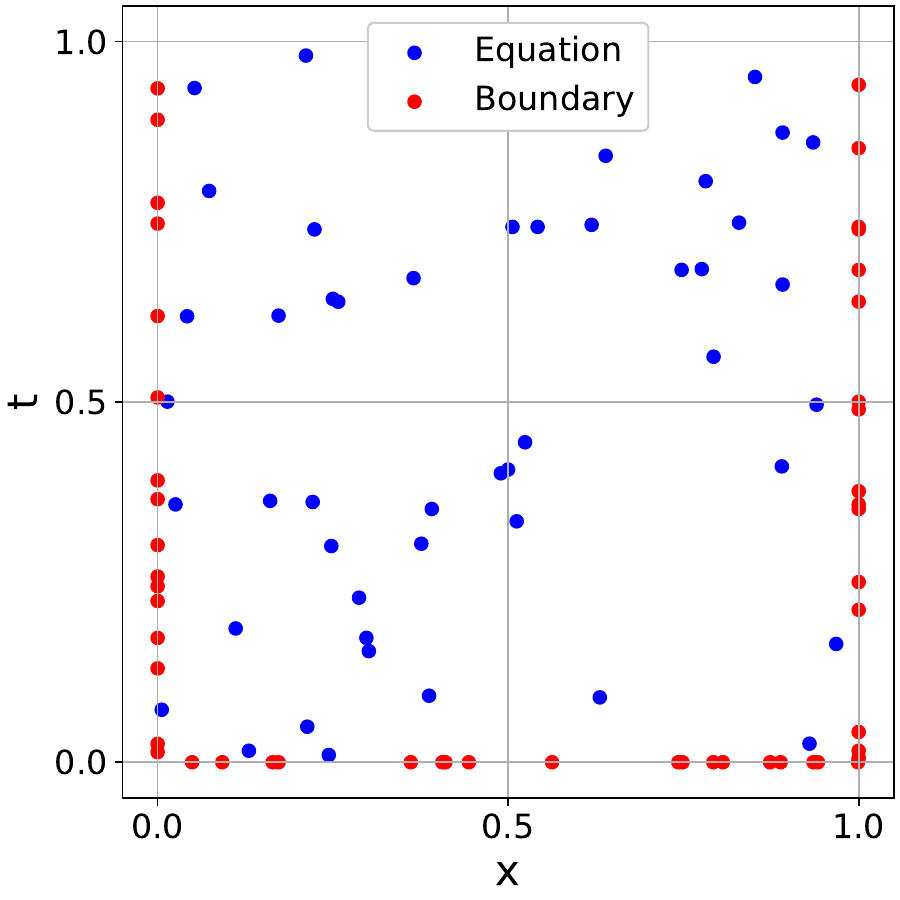}\\
	\includegraphics[width=0.4\textwidth]{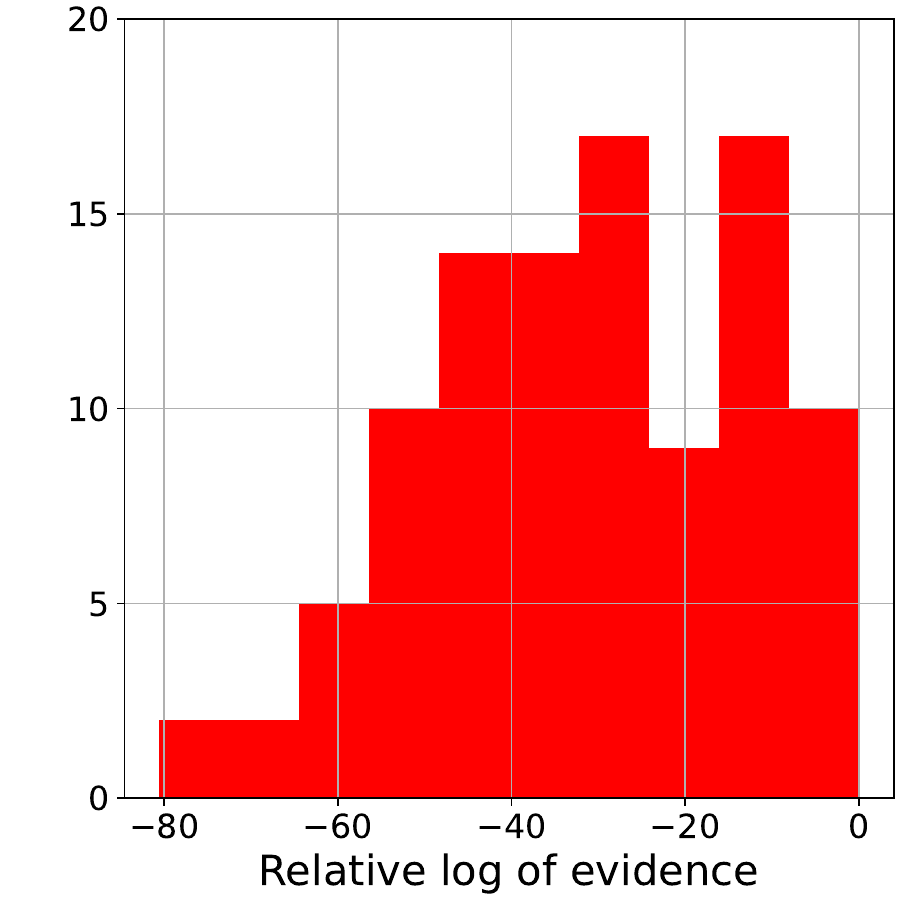}
    \end{center}
	\caption{In the top panel: the training data, the blue/red points correspond to the equation/boundary contribution to the loss for the heat equation. In the bottom panel: the histogram of the relative log of evidence (the difference between a given log of evidence and the highest log of evidence).}
	\label{fig:HeatEquationPoints_evidence}
\end{figure}    

The data on the boundary Eqs. (\ref{Eq:Heat_b1}-\ref{Eq:Heat_b2}) were randomly drawn from the uniform distribution ($50$ points). Similarly, we obtained data in $(0,1)\times (0,1)$ (50 points). The data considered in the training are shown in Fig. \ref{fig:HeatEquationPoints_evidence} (left panel).

%

Here, the Bayesian training was performed in the simplest fashion. Namely, we kept one $\alpha$ parameter for all weight classes (as in Ref. \cite{Graczyk:2010gw})  and one $\beta = \beta_q=\beta_b$ for equation and boundary contribution. 
\begin{figure*}[h]
	\centering
	\includegraphics[width=\textwidth]{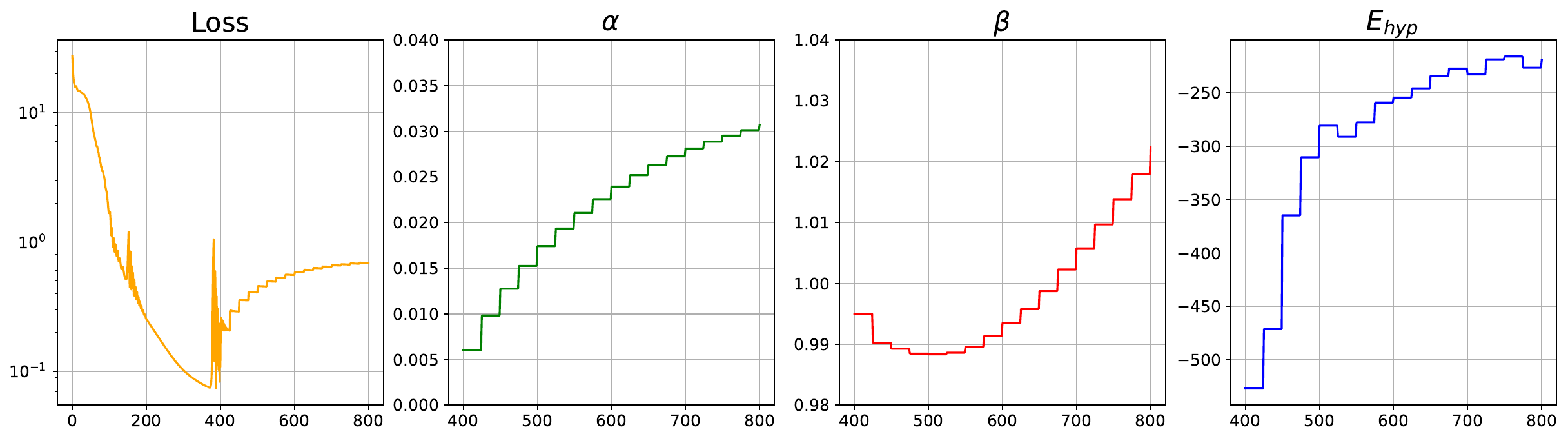}
	\caption{An example of loss $E_T$ (the first panel on the left), $E_{hyp}$ (the last panel), and $\alpha$ (the second from the right) and $\beta$ (the third from the right) hyperparameters evolution during the training of the network that solves the heat equation. 
 }
	\label{fig:HeatEquationTraining}
\end{figure*}

We optimized the hyperparameters after $5000$ epochs to maintain the procedure's convergence. Then, every $25$ epoch, we updated the $\alpha$ and $\beta$ parameters to minimize the error $E_{hyp}$. In Fig.~\ref{fig:HeatEquationTraining}, we show one of the registered training examples. The method leads to convergence of both discussed losses, $E_T$ and $E_{hyp}$, with respect to weights and hyperparameters, respectively. We started the optimization with a low value of $\alpha$ (the penalty term was of minor relevance). During the training, $\alpha$ changes by the order of magnitude. The $\beta$ parameter slowly varies during the optimization process.

\begin{figure*}[h]
	\centering
	\includegraphics[width=\textwidth]{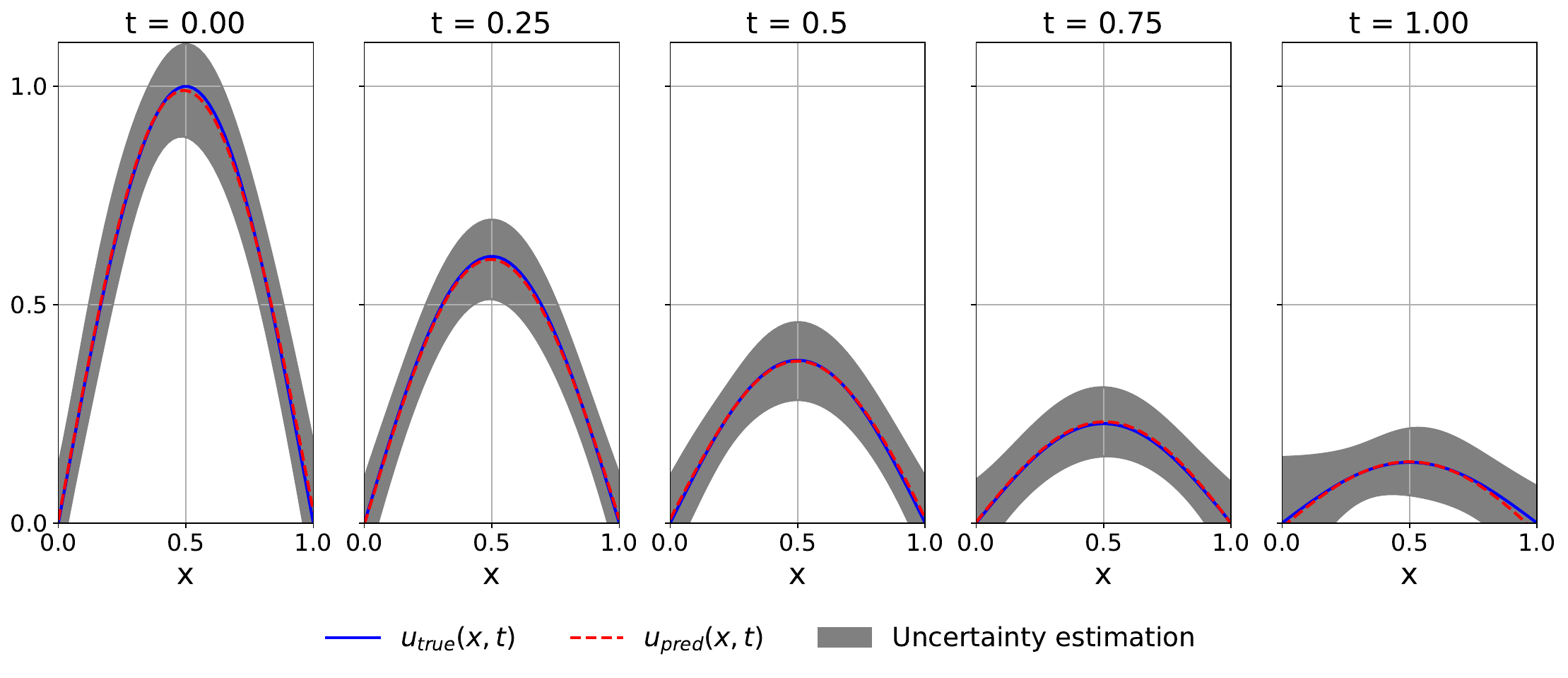}
	\caption{The exact ($u_{true}$) and PINN ($u_{pred}$) solutions for the heat equation. The grey area  denotes $2\sigma$ uncertainty. \label{fig:HeatEquationSolution}}
\end{figure*}

For every fit, we compute the evidence. In Fig.~\ref{fig:HeatEquationPoints_evidence} (right panel), we show the histogram of evidence values for $100$ models. The best model has the highest value of evidence. In Fig.~\ref{fig:HeatEquationSolution} we show our best fit together with the exact analytic solution. The fit is plotted together with $2\sigma$ uncertainty. Note that we show the $x$-dependence of the solution in five time steps.   

\subsection{Wave equation}

\label{Sec:NumRes:wave}

Let us consider the following wave equation
\begin{eqnarray}
\label{Eq:Wave}
\frac{\partial^2 u}{\partial t^2} -  \frac{\partial^2 u}{\partial x^2} &=& 0, \quad x \in [0,1],\; t \in [0, 1]
\\
\label{Eq:Wave_b1}
u(0, t) = u(1, t) &=& 0
\\
\label{Eq:Wave_b11}
\frac{\partial}{\partial t} u(0, t) &=& 0
\end{eqnarray}
with
\begin{equation}
\label{Eq:Wave_b2}
u(x, 0) = \sin(\pi x) + \frac{1}{2}\sin(2\pi x).
\end{equation}
The analytic solution reads 
\begin{equation}
u_{\text{true}}(x, t) = \sin(\pi x) \cos( \pi t) + \frac{1}{2}\sin(2\pi x) \cos(2 \pi t).
\end{equation}

\begin{figure}[h]
	{\centering
	\includegraphics[width=0.4\textwidth]{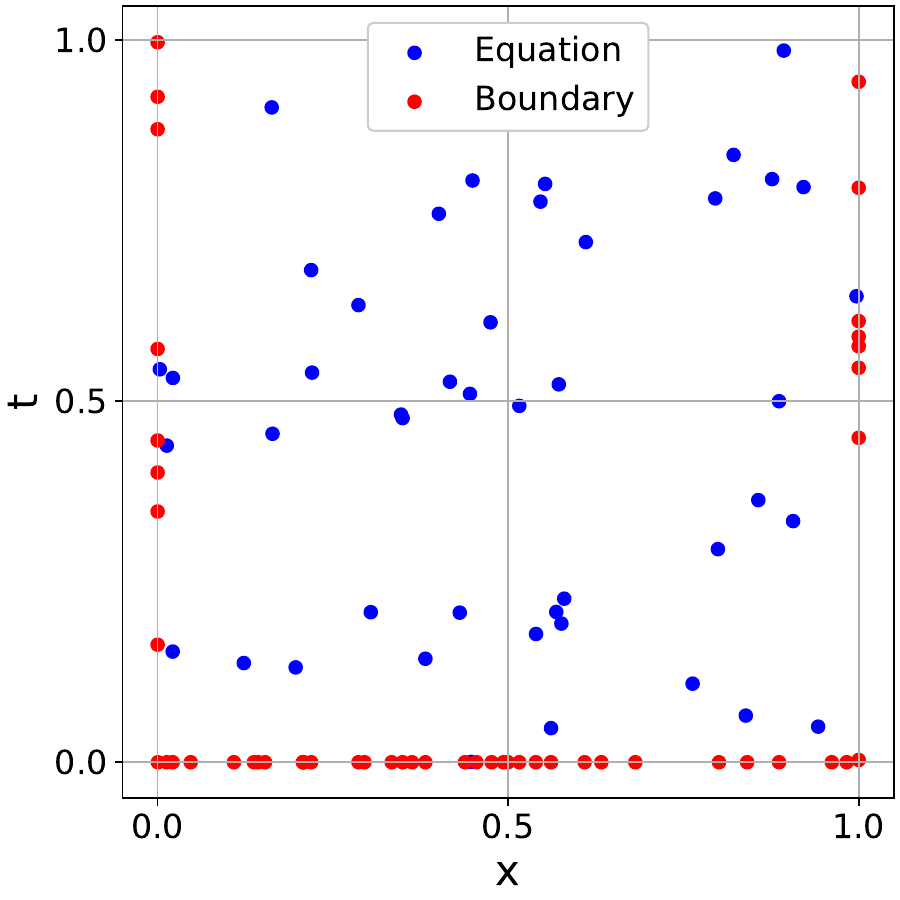}\\
	\includegraphics[width=0.4\textwidth]{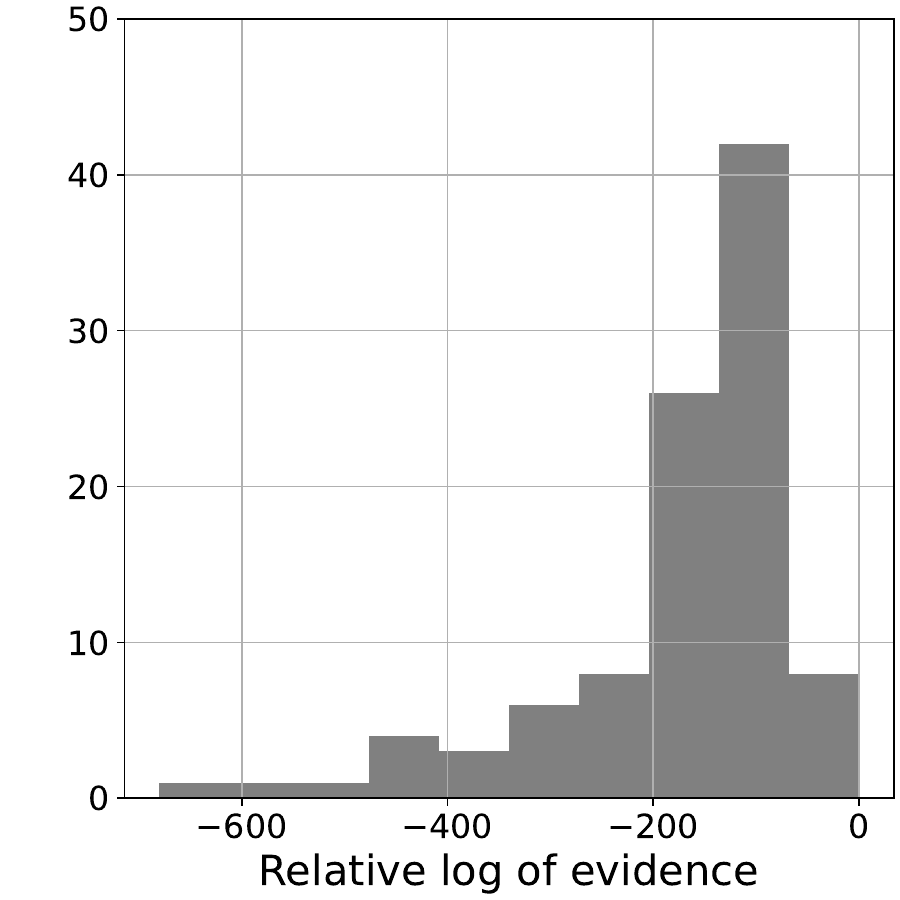}}
	\caption{In the top panel: the training data, the blue/red points correspond to equation/boundary contribution to the loss for the wave equation. In the bottom panel: the histogram of the relative log of evidence (the difference between a given log of evidence and the highest log of evidence).}
	\label{fig:WaveEquationPoints}
\end{figure}

In the previous subsection, we demonstrated that the BF is effective for solving the heat equation. The solution was obtained using a relatively small network, and the optimization process took only a few epochs. However, solving the wave equation using BF proved more challenging. We used a network with one hidden layer containing 8 hidden units, with the hyperbolic tangent as the activation function. To get a successful fit, we generated a dataset of the same size as the heat equation problem. The dataset consisted of $50$ points for $\mathcal{M}$ and 50 points for $\partial\mathcal{M}$, drawn from a uniform distribution. We also considered more points for the initial condition at $t=0$. The dataset's distribution is shown in the left panel of Fig.~\ref{fig:WaveEquationPoints}.

%
%

\begin{figure*}[h]
	\centering
	\includegraphics[width=\textwidth]{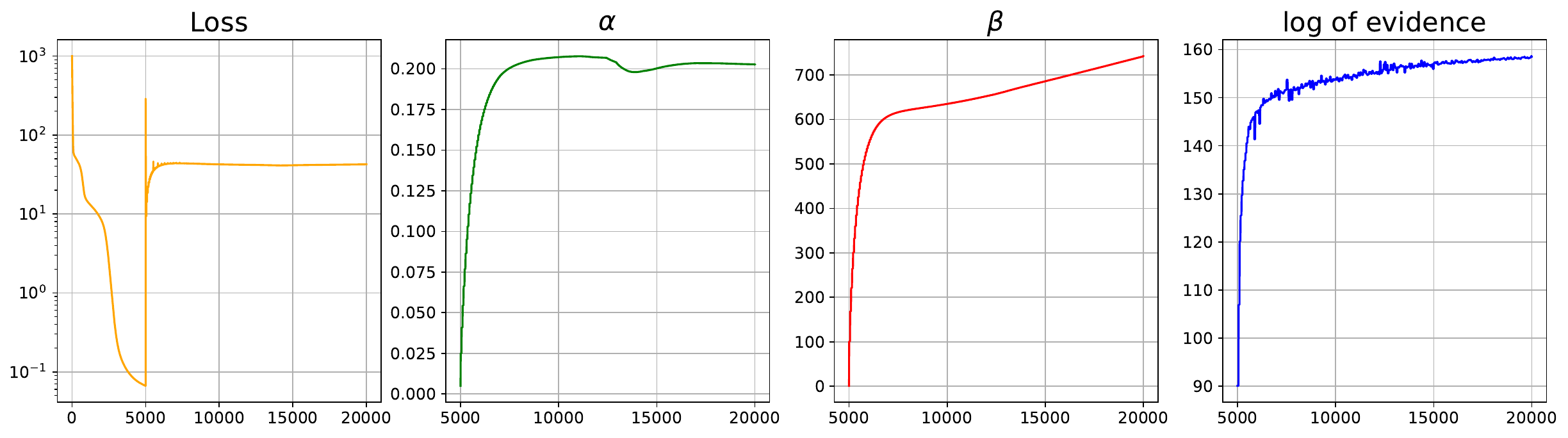}
	\caption{Parameters evolution during the wave equation training for the model with the highest value of evidence.}
	\label{fig:WaveEquationTraining}
\end{figure*}

\begin{figure*}[h]
	{\centering
		\includegraphics[width=\textwidth]{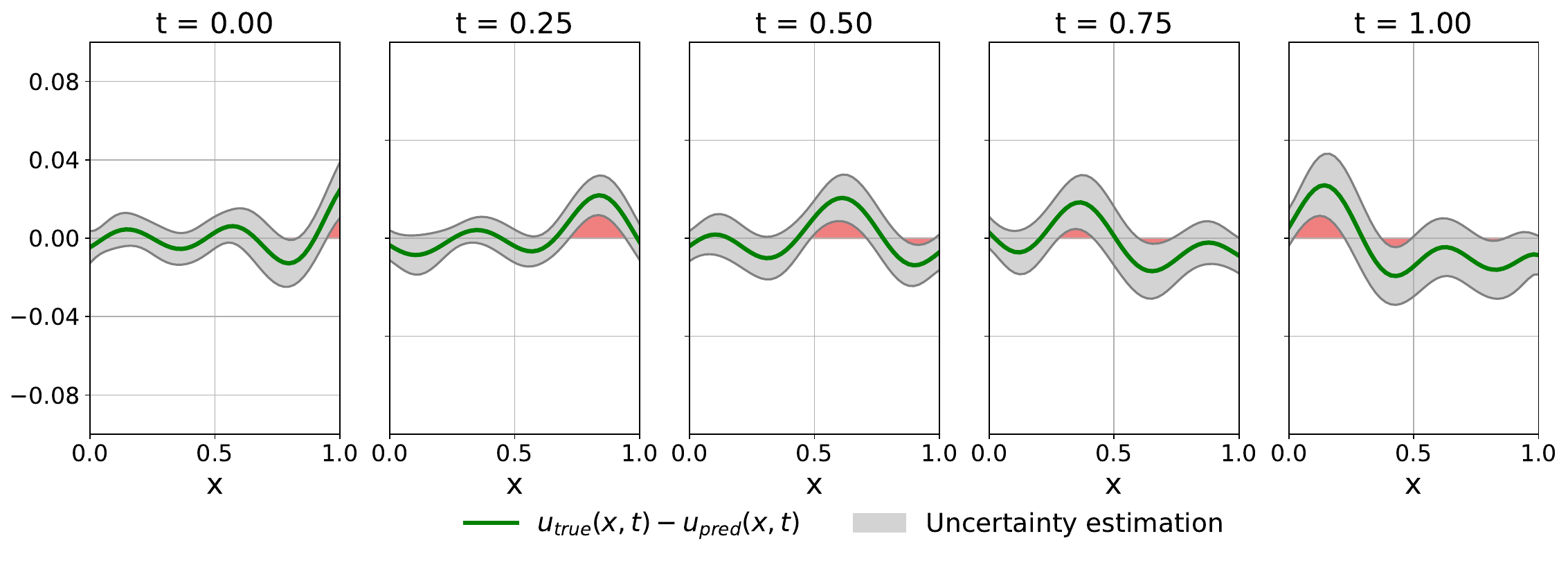}}
	\caption{Comparison of the obtained uncertainties and real error for the solution of the wave equation with boundary condition (\ref{Eq:Wave_b2}). The $y=0$ axes correspond to the exact agreement between true. The red color indicates were true values are outside the $2\sigma$ uncertainty interval.}
	\label{fig:WaveEquationError}
\end{figure*}

The wave equation with boundary condition (\ref{Eq:Wave_b2}) was solved in the simplified framework with only one $\alpha$ and $\beta$ hyperparameters. In this case, the learning process lasted significantly longer than in the heat equation problem, requiring 20,000 epochs. We started optimization of the hyperparameters in the $5000$th epoch. The best (due to evidence) training example for this case is given in Fig.~\ref{fig:WaveEquationTraining}. 
\begin{figure*}[h]
	\centering
	\includegraphics[width=\textwidth]{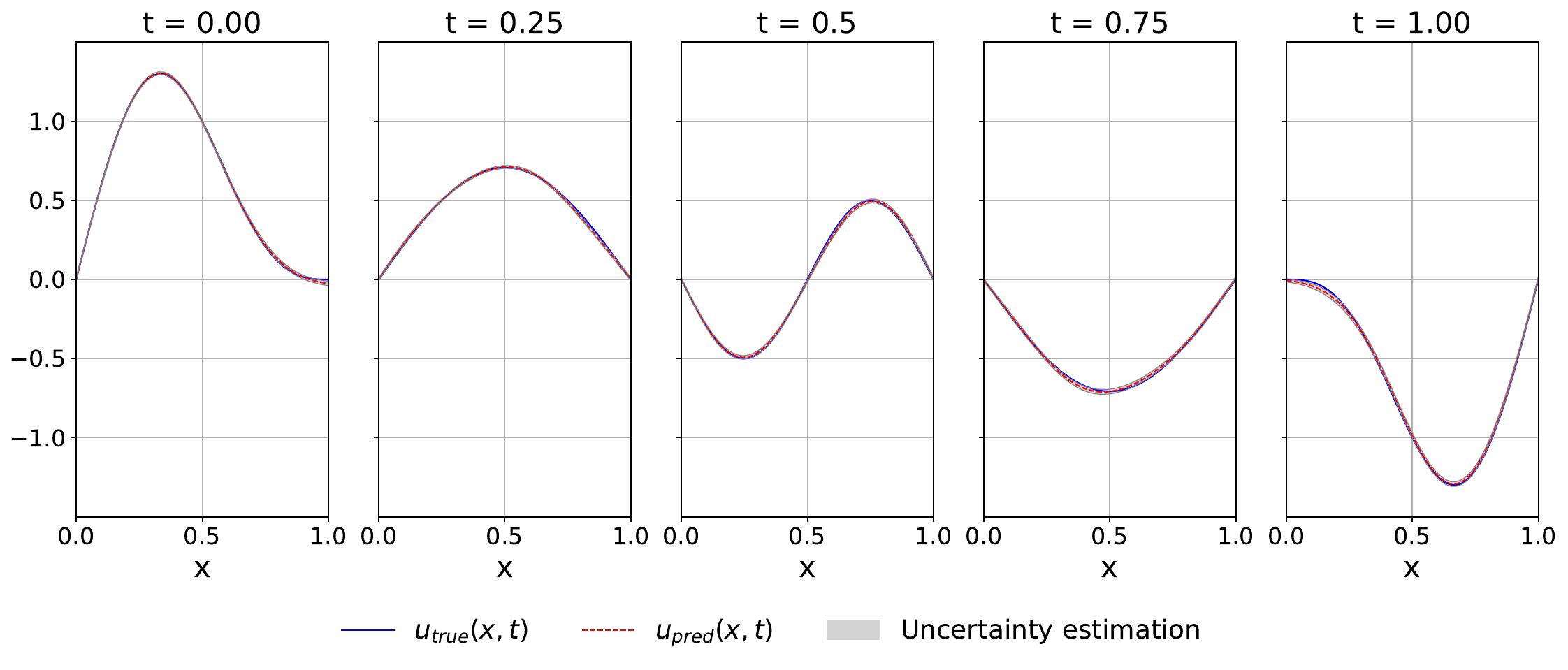}
	\caption{The exact ($u_{true}$) and PINN ($u_{pred}$) solutions  for the wave equation with boundary conditions (\ref{Eq:Wave_b2}). The grey area  denotes $2\sigma$ uncertainty. }
	\label{fig:WaveEquationSolution}
\end{figure*}
Similarly, as for the heat equation, we show the histogram of the relative log of evidence values in 
the right panel of Fig.~\ref{fig:WaveEquationPoints}. Our best fit agrees with the true solution; see Fig.~\ref{fig:WaveEquationSolution}. However, one can notice a slight difference between the wave equation's exact and network model solutions. To visualize this feature, we plot $u_{true}(x,t)-u_{pred}(x,t)$ together with uncertainty in Fig.~\ref{fig:WaveEquationError}.  

\begin{figure*}[h]
	\centering
	\includegraphics[width=\textwidth]{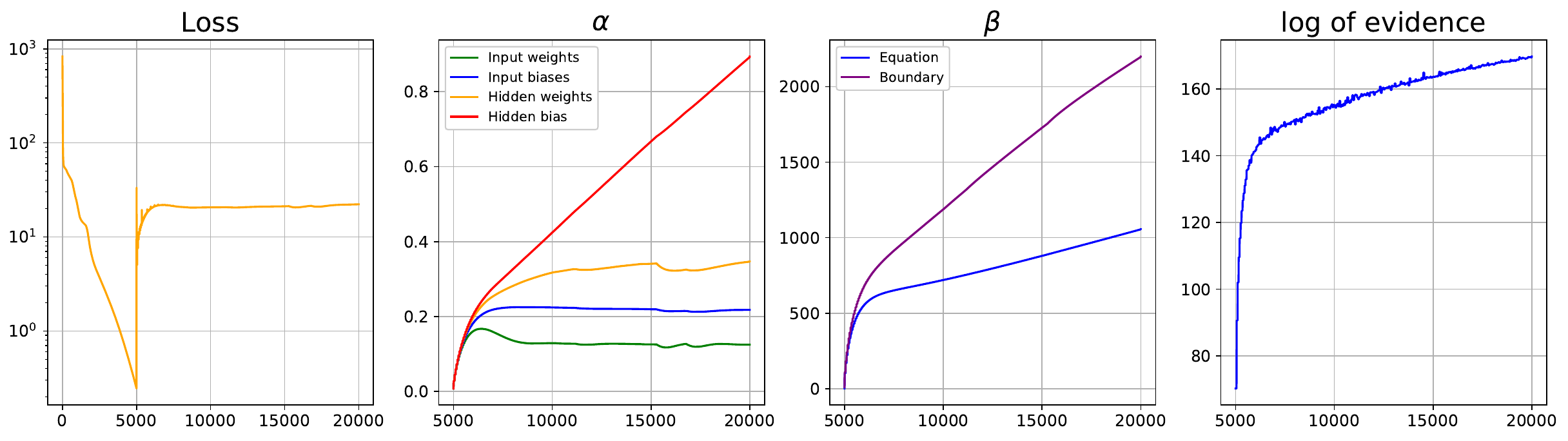}
	\caption{The parameters evolution for the full model. The loss function and evidence reached their plateaus. Although, as one can see, the convergence of all parameters was not obtained in the training process.}
	\label{fig:WaveEquationAlphaBetaTraining}
\end{figure*}

Having more than one $\alpha$ and $\beta$ parameter complicates numerical computation, making it harder to achieve convergence. However, we performed a complete analysis of the same equation, the dataset, and the training parameters, keeping for every layer of units weights one $\alpha$ and a separate $\alpha$ for bias units. We also hold two betas, namely $\beta_q$ for the equation and $\beta_b$ for the boundary contribution. Hence, in the analysis, we have four $\alpha$ parameters (two for the input layer and two for the hidden layer: one for ordinary weights and one for bias weights), and two betas. Again, we considered $100$ models in the analysis, and the best-fit evidence (log-likelihood) was about $169$. In training with one $\alpha$ and one $\beta$, the best model had an evidence (log-likelihood) of about $158$, indicating that the model with fully allowed hyperparameters is more favorable under the Bayesian algorithm. The evolution of the hyperparameters during training is shown in Fig. \ref{fig:WaveEquationAlphaBetaTraining}. Optimizing the model leads to convergence of the loss and the log-evidence. When one approaches the minimum of the loss, the boundary constraints' importance becomes increasingly important; in fact, $\beta_b$ is almost twice as large as $\beta_q$, when the number of data points from both contributions is the same.

To summarize, the approach used here is quite effective. However, in more complex numerical applications, we'll simplify by considering only a single regularizer, denoted by $\alpha$, and a complete set of beta parameters. In this scenario, we can effectively regulate the model by tuning all the hyperparameters that describe the equation and the boundary constraints using the Bayesian algorithm.

\subsection{Burger's equation}

\label{Sec:NumRes:Burgers}

For the last example of PDEs, we consider Burger's equation with Dirichlet boundary conditions.
\begin{eqnarray}
\frac{\partial u}{\partial t} + u \, \frac{\partial u}{\partial x} - \frac{1}{100 \pi}
\frac{\partial^2 u}{\partial x^2} &=& 0,    
\label{Eq:Burger}
\\
& &  \quad x \in [-1,1], \;\; t\in[0,1] \nonumber
\\
u(0,x) &=& - \sin (\pi x)
\label{Eq:Burger_b1}
\\
u(t,-1) &=& u(t,1) = 0.
\label{Eq:Burger_b2}
\end{eqnarray}
It is the same equation as discussed in Ref.~\cite{raissi2017physicsI}.
\begin{figure*}[h]
	\centering
	\includegraphics[width=0.4\textwidth]{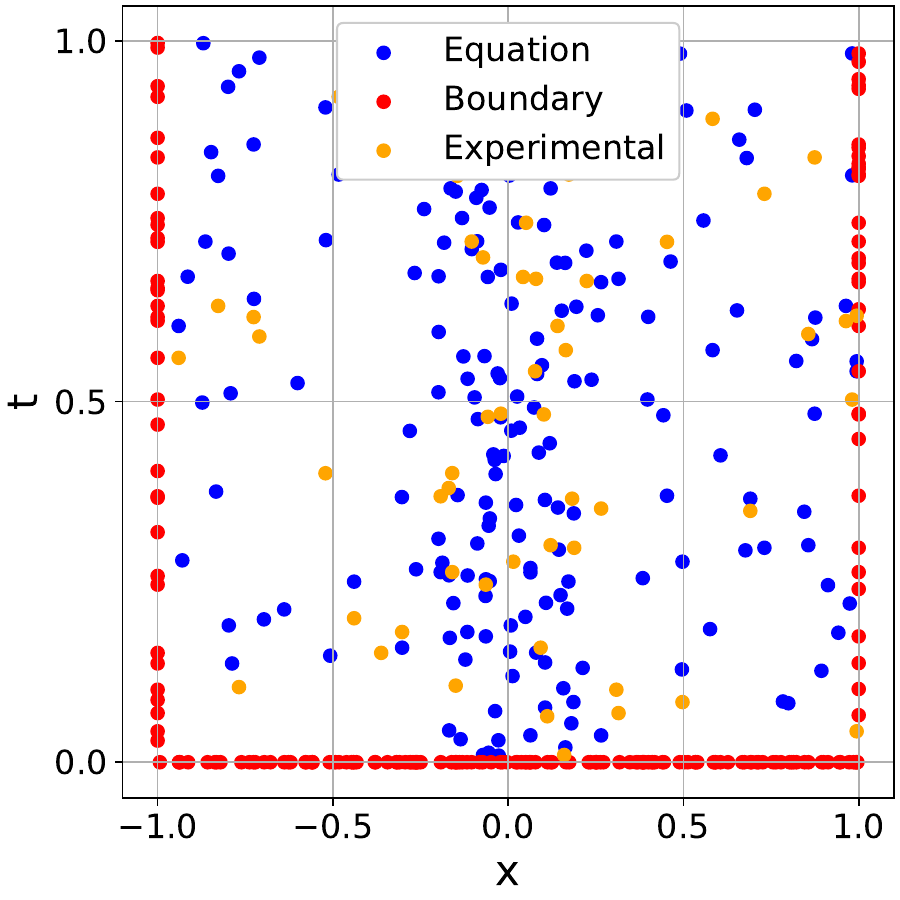}
	\includegraphics[width=0.4\textwidth]{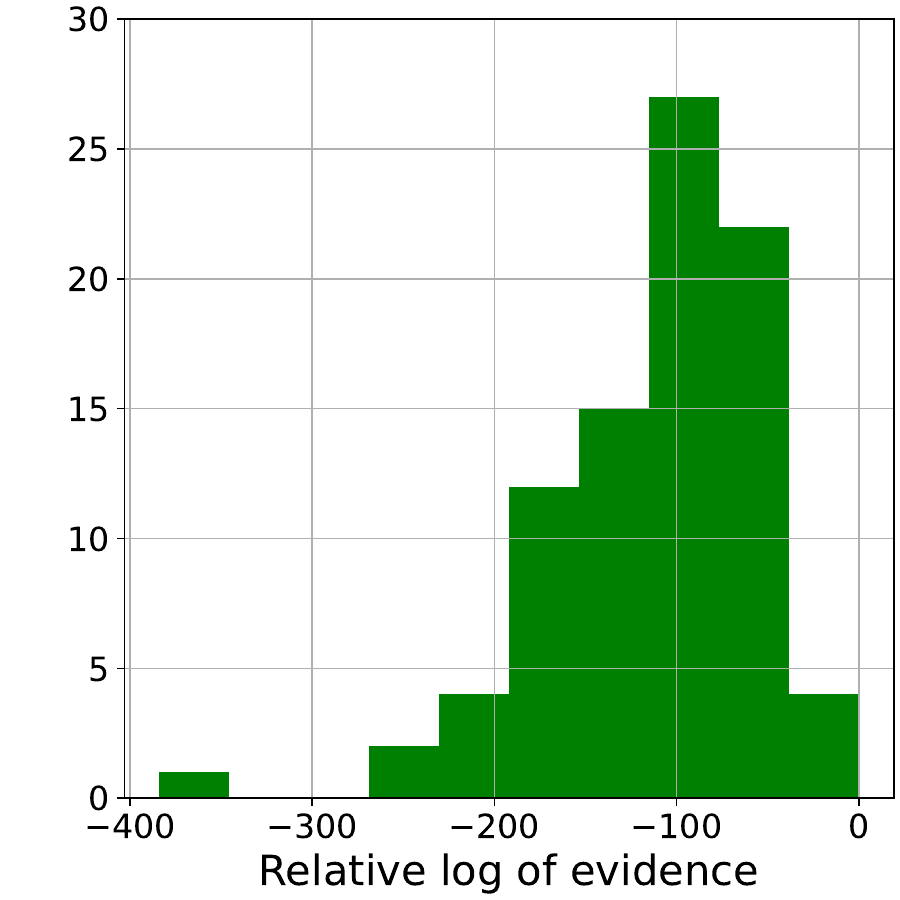} 
	\caption{In the left panel: the training data, the blue/red points correspond to the equation/boundary contribution to the loss for Burger's equation. In the right panel: the histogram of the relative log of evidence (the difference between a given and the highest log of evidence). \label{fig:BurgerEquationPoints}
	}
\end{figure*}
\begin{figure*}[h]
	\centering
	\includegraphics[width=\textwidth]{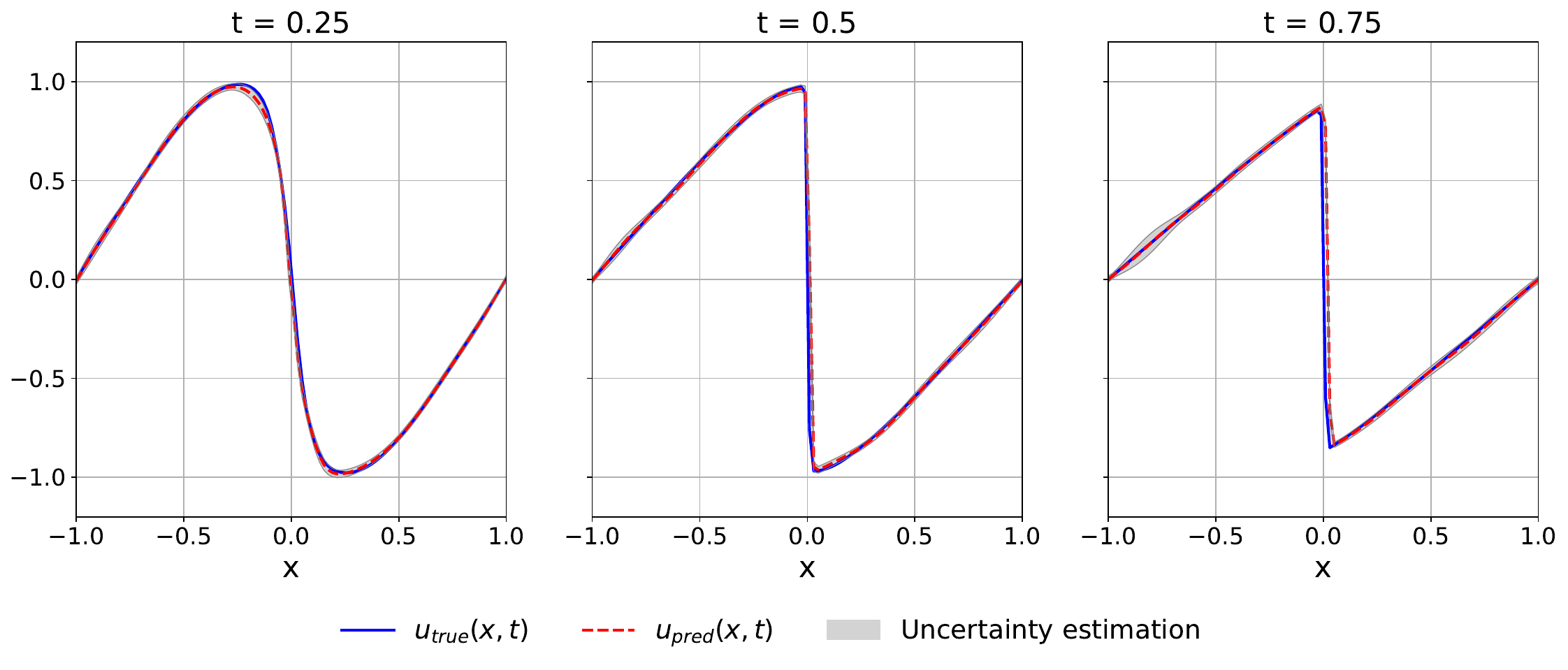}
	\caption{The ''true'' solution ($u_{true}$) - the numerical solution from \cite{raissi2017physicsI}, and our best model predictions ($u_{pred}$) for Burger's equation. The grey area  denotes $2\sigma$ uncertainty. }
	\label{fig:BurgerEquationSolution}
\end{figure*}

Burger's equation is the most difficult to solve among the three PDEs discussed in this paper. Since the solution's time evolution goes from smooth sine to a very sharp function, solving the problem with one hidden layer is difficult, so we employed a neural network architecture with two hidden layers with eight units each and, again, hyperbolic tangent as the activation function.

For this problem, we adopt a more general approach that includes additional data points. In this case, the loss function is as follows
\begin{align}
    E_T &= \sum_{i=1}^C \alpha_i E_w^i + \beta_{q} E_D^{q} + \beta_{b} E_D^{b} + \beta_{ex} E_D^{ex}.
\end{align}

The dataset contained $200$, $200$, and $60$ points for equation, boundary, and experimental data, respectively. The noise in the ``experimental`` data was equal to 0.02. We sampled half of the ``experimental`` and equation points from the uniform distribution in the interval $x\in[-0.2, 0.2]$ and the rest from $x\in[-1, 1]$ ($t$ variable is drawn from uniform distribution in $[0,1]$). Obtaining an accurate solution with limited data was possible when there were enough points near $x=0$. We show the data distribution in Fig.~\ref{fig:BurgerEquationPoints} (left panel). 

The networks were trained for $20000$ epochs, and we started fitting parameters since the $10000$th epoch.  In this case we used one $\alpha$ for all weights, but we optimized $\beta_q$, $\beta_b$, $\beta_{ex}$ separately. 

Similarly, we collected $100$ fits for two previous PDE examples. The histogram of the log of evidence for collected fits is given in Fig.~\ref{fig:BurgerEquationPoints} (right panel). In Fig.~\ref{fig:BurgerEquationSolution}, we show our best fit with the numerical solution obtained in Ref.~\cite{raissi2017physicsI}. A good agreement between the predictions of both approaches is seen.

\subsection{Numerical efficiency}

\begin{table}[ht]
\centering
\caption{\raggedright{Summary of evaluations. L2 (mean square error) computed on 10,000 grid points for the best model in the ensemble of $100$ networks. Average time of getting the model.}}
\label{Table_evaluation_summary}
\begin{ruledtabular}
\begin{tabular}{|c|c|c|}
\hline\hline
Problem       & L2     & time [s] per fit \\ 
\hline
Heat Equation & 0.051 & 59   \\ 
Wave Equation & 0.119 & 152  \\ 
Wave equation (many $\alpha's$) & 0.011 & 159 \\ 
Burger Equation & 0.103 & 588 \\
\hline
\end{tabular}
\end{ruledtabular}
\end{table}

Table~\ref{Table_evaluation_summary} summarizes the predictive accuracy and computational cost of the evidence-selected models for each PDE. The reported L2 (mean squared error) values correspond to the best model (i.e., the model with the strongest evidence) selected from an ensemble of $100$ independently trained networks. The $L_2$ error was evaluated on a grid of 10\,000 points. The table also reports the average wall-clock time required to obtain a single trained model.

For the heat equation, a small network with a single decay hyperparameter achieves accurate solutions at modest computational cost. The wave equation is more demanding due to its oscillatory structure and its second-order time derivative, leading to larger errors and longer training times. Allowing multiple decay hyperparameters significantly improves accuracy, at the cost of a moderate increase in computational cost.

The Burgers equation represents the most challenging case considered. Owing to the nonlinear convective term and the development of sharp solution features, deeper architectures and longer optimization are required, which leads to the highest computational time per fit. Nevertheless, the evidence-based selection procedure consistently identifies stable and accurate solutions across all investigated problems.

To assess the practical impact of the evidence-based hyperparameter optimization, we compared the proposed Bayesian PINN with a standard PINN using fixed hyperparameters $\alpha$ and $\beta$. Both approaches employed identical network architectures, training datasets, and optimization settings. The only difference was that in the baseline model, the loss weights were kept constant throughout training. Indeed, we keep the hyperparameters fixed at $\alpha=0.05$ and $\beta=1.0$.

In Table~\ref{Table_PINN_vs_BPINN}, we present the mean L2 over the five best results according to the PINN (the lowest error) and our approach (the strongest evidence). For each case, we provide the mean absolute deviation (MAD).

For the heat equation, both approaches achieved comparable accuracy, indicating that fixed loss weights are sufficient for relatively simple problems in which the relative importance of the residual and boundary terms is naturally balanced.

In contrast, for the wave equation, the Bayesian PINN consistently achieved lower errors across multiple independent runs. Averaged over five trainings, the Bayesian approach reduced the relative $L_2$ error by approximately $25\%$ compared to the fixed-weight baseline. A similar reduction was observed for MAD. These results demonstrate that evidence-driven tuning of loss weights becomes particularly beneficial for more challenging PDEs, where the balance between equation and boundary contributions is nontrivial.

\begin{table}[h]
\centering
\caption{Comparison between fixed-weight PINN and Bayesian PINN (our approach).
Reported values are averages over five independent runs, except L2$_{best}$, which is the best according to each framework result. \label{Table_PINN_vs_BPINN}}
\begin{ruledtabular}
\begin{tabular}{|l|c|c|c|c|}
\hline
Problem & Method &  L2$_{best}$ &  $\overline{\mathrm{L2}}$ & $\overline{\mathrm{MAD}}$ \\
\hline
Heat Equation & Fixed weights & 0.00538 & 0.00588 & 0.00446 \\
Heat Equation & Bayesian & 0.00508 & 0.00678 & 0.00530 \\
\hline
Wave Equation & Fixed weights & 0.0129 & 0.01514 & 0.01248 \\
Wave Equation & Bayesian & 0.01146 & 0.01114 & 0.00903 \\
\hline
\end{tabular}
\end{ruledtabular}
\end{table}

\subsection{Comparison with other Bayesian approaches}

Depending on the method used to evaluate posterior densities, one may distinguish at least three types of Bayesian approaches. 
In the first class, the Laplace approximation is applied (as in the present work). 
In the second class, variational methods are employed; an example is given in Ref.~\cite{Zhu_2018}, where the authors adopt a non-parametric variational inference method known as stochastic variational gradient descent (SVGD)~\cite{liu2016svgd}. 
The third class comprises Monte Carlo-based methods, such as the Hamiltonian Monte Carlo approach discussed in a previous section.

The Laplace-based method is computationally efficient for relatively small and shallow neural networks. 
Variational approaches are generally more scalable and can be applied to larger network architectures. 
In contrast, Monte Carlo-based methods typically require substantial computational resources and are therefore better suited to smaller models. 
A more detailed comparison between our approach and the remaining two methods is presented in Table~\ref{Table_Bayes}.

Summarizing Bayesian deep learning approaches, such as \cite{Zhu_2018}, employ variational inference to approximate posterior distributions over NN parameters. While these methods can scale to larger architectures, they typically require stochastic gradient optimization of variational distributions and repeated sampling. In contrast, the evidence-based approach adopted here uses a Laplace approximation around the maximum posterior solution, enabling direct computation of model evidence and efficient hyperparameter tuning. This makes the method particularly attractive for moderately sized PINNs where Hessian-based inference is computationally feasible.

Bayesian deep learning approaches have also been applied to surrogate modeling of PDE dynamics. For example, Geneva and Zabaras \cite{GENEVA2020109056} proposed a physics-constrained auto-regressive neural network with Bayesian uncertainty estimation for predicting the time evolution of PDE systems. Unlike PINN-based approaches, their method learns a surrogate mapping between solution states rather than directly solving the governing equations through residual minimization.

\begin{table*}[ht]
\centering
\caption{\raggedright{Summary of three Bayesian approaches: the evidence-based Laplace approximation (this work), variational inference, and Hamiltonian Monte Carlo (HMC).}}
\label{Table_Bayes}
\begin{ruledtabular}
\begin{tabular}{|l|l|l|l|}
\hline
\textbf{Aspect} & \textbf{This work} & \textbf{Variational} & \textbf{HMC-based} \\
\hline
Posterior approximation & Laplace approximation & Variational approximation & Sampling \\
Computational cost & Low & Medium & High \\
Hyperparameter tuning & Evidence-based & Variational / heuristic & Manual \\
Scalability & Shallow PINNs & Moderate to large networks & Limited (small models) \\
\hline
\end{tabular}
\end{ruledtabular}
\end{table*}

\section{Summary}

\label{Sec:Summary}

We introduced the Bayesian framework for the PINN to solve heat, wave, and Burger's equations. The network solutions agree with the "true" ones. The method allowed us to compute the  uncertainty due to variation in network parameters, and one may also estimate the variation due to changes in the network architecture. The latter information is encoded in the evidence distribution. Hence, the presented method allows us to estimate epistemic uncertainties. 

We have shown that our approach can tune relative weights among the various contributions to the loss function. Our method differs from those discussed in Sec.\ref{Sec:Adaptive_losses_review}. In our approach, the relative weights, treated as hyperparameters, are tuned separately for each loss component. Indeed, to optimize $\beta$s, we minimize $E_{hyp}$, which depends partially linearly on them. But, there is a nonlinear dependence, too, which is hidden in the $\ln |\mathbf{A}|$ and logarithmic contributions. The last two types contribute to the so-called Occam factor and penalize a model formulation that is too complex. Hence, the Bayesian algorithm ''chooses'' the optimal configuration of conditions according to the delivered information. Note that, in principle, in the presented Bayesian approach, one would introduce a corresponding hyperparameter for each data point, but studying this problem goes beyond the scope of this paper. 

In the present manuscript, we consider shallow neural networks of small sizes. For such architectures, the numerical procedure is effective. Computing the Hessian is accurate and takes little time. Moreover, as mentioned, the Gaussian approximation makes sense when the number of model parameters is smaller than the number of constraints~\cite{MacKay1992.4.3.448}. 

Summarizing: the method is intended for shallow or moderately sized PINNs, where Hessian-based inference is computationally feasible; extension to deep architectures would require alternative posterior approximations.

\newpage

\appendix

\section{General foundations of BF}
\label{Appendix_0}

The Bayesian framework we consider aims to obtain the posterior densities for the network parameters, weights $\mathbf{w}$, and hyperparameters $\boldsymbol \alpha$ and $\boldsymbol \beta$.  

A choice of network architecture impacts the analysis results. Indeed, a model that is too simple, with a small number of neurons, cannot be flexible enough to describe the problem. On the other hand, a model that is too complex, with many neurons, may tend to overfit the data and, as a result, not interpolate well between grid points. As mentioned above, BF shall allow us to compare fits across different networks and obtain posterior densities for all necessary model parameters and hyperparameters. 

In the BF the posteriors for the model parameters and hyperparameters are obtained in the ladder approximation described below. Finally, the approach leads to the computation of the so-called evidence, a measure that ranks models.

The starting point of the procedure is to obtain the posterior density for the network weights. Assuming that the hyperparameters, $\boldsymbol \alpha$ and $\boldsymbol \beta$, and model architecture $\mathcal{N}$ are fixed
\begin{eqnarray}
P(\mathbf{w} \mid \boldsymbol \alpha , \boldsymbol \beta, \mathcal{N}, \mathcal{D}) 
&=& \frac{P(\mathcal{D}  \mid \mathbf{w},  \boldsymbol \beta, \mathcal{N}) P(\mathbf{w} \mid \boldsymbol \alpha, \mathcal{N}) }{ P(\mathcal{D}  \mid  \boldsymbol \alpha , \boldsymbol \beta, \mathcal{N}) }
\label{Eq:Posterior_w_general_ladder1}
\\
P(\mathcal{D}  \mid  \boldsymbol \alpha , \boldsymbol \beta, \mathcal{N}) & = &
\int  P(\mathcal{D}  \mid \mathbf{w},  \boldsymbol \beta, \mathcal{N}) P(\mathbf{w} \mid \boldsymbol \alpha, \mathcal{N}) d \mathbf{w}. \nonumber \\
\label{Eq:Posterior_w_general_ladder2}
\end{eqnarray}

In the above, we silently assumed that the hyperparameter $\boldsymbol \beta$ appears in the likelihood \\ $P(\mathcal{D}\mid\mathbf{w},\boldsymbol\beta,\mathcal{N})$, while hyperparameter $\boldsymbol \alpha$ appears in prior $P(\mathbf{w} \mid \boldsymbol \alpha, \mathcal{N})$ only.

In principle, the hyperparameters should be integrated from the model's prior densities, but usually, it is either difficult or impossible to perform. In the BF framework, it is assumed that the posterior density $p(\boldsymbol \alpha, \boldsymbol \beta \mid \mathcal{N}, \mathcal{D} )$ 
has a sharp peak at $\boldsymbol{\alpha}_{MP}$, $\boldsymbol{\beta}_{MP}$ ($MP$ - maximum of posterior) then prior for weights reads $p(\mathbf{w} \mid \mathcal{N},\mathcal{D} ) \approx p(\mathbf{w} \mid \boldsymbol{\alpha}_{MP},\boldsymbol{\beta}_{MP},\mathcal{N},\mathcal{D} )$ (in reality, the posterior has many sharps peaks, we comment on that in the following section).  We see that to obtain the desired prior, one needs to find $\boldsymbol{\alpha}_{MP}$, $\boldsymbol{\beta}_{MP}$.

In the second step, we use the likelihood for the hyperparameters, Eq.~\ref{Eq:Posterior_w_general_ladder2}, to evaluate the posterior density for the hyperparameters, namely,
\begin{eqnarray}
P(    \boldsymbol \alpha , \boldsymbol \beta\mid \mathcal{N},\mathcal{D}) &= & \frac{ P(\mathcal{D}  \mid  \boldsymbol \alpha , \boldsymbol \beta, \mathcal{N}) P(\boldsymbol \alpha, \boldsymbol \beta \mid \mathcal{N})}{P(\mathcal{D}  \mid  \mathcal{N})}
\label{Eq:Posterior_hyper_general_ladder1}
\\
\label{Eq:posterior_alpha_beta}
P(\mathcal{D}  \mid  \mathcal{N}) &=& \int  P(\mathcal{D}  \mid  \boldsymbol \alpha , \boldsymbol \beta, \mathcal{N}) P(\boldsymbol \alpha, \boldsymbol \beta \mid \mathcal{N}) d \boldsymbol \alpha \,d \boldsymbol \beta \nonumber \\
\label{Eq:Posterior_hyper_general_ladder2}
\end{eqnarray}
Note that the density (\ref{Eq:Posterior_hyper_general_ladder2}) is called the model evidence. In reality, the prior for the hyperparameters factorizes into the product of two densities, namely,
\begin{equation}
P(\boldsymbol \alpha, \boldsymbol \beta \mid \mathcal{N}) 
= P(\boldsymbol \alpha \mid \mathcal{N})\, P(\boldsymbol \beta \mid \mathcal{N}).
\end{equation}

In the third step, we compute the posterior density $P(\mathcal{N}\mid \mathcal{D})$. It is the quantity that ranks the models, i.e., a model the most favorable by the data has the highest value of $P(\mathcal{N}\mid\mathcal{D})$.
\begin{eqnarray}
\label{Eq:Posterior_evidence_general_ladder1}
P(\mathcal{N}\mid \mathcal{D})  
&=& \frac{P( \mathcal{D} \mid \mathcal{N}) P(\mathcal{N}) }{P(\mathcal{D})}.
\end{eqnarray}
In practice, any model is equally likely at the beginning of the analysis, and $P(\mathcal{N})$ is uninformative. Hence
\begin{eqnarray}
\label{Eq:Posterior_evidence_general_ladder1b}
P(\mathcal{N}\mid \mathcal{D})  
& \sim & P( \mathcal{D} \mid \mathcal{N}) 
\end{eqnarray}
and to rank the model, it is enough to compute the evidence.

\section{Three steps of Bayesian inference}
\label{Appendix_A}
This section summarizes the main steps of our framework. The main idea is to assume that all the densities should have a Gaussian-like shape. The inference process of estimating optimal weights, hyperparameters, and model structure can be split into three steps announced in the previous section.

\subsection{First step}

The prior density for the weights is constructed based on several observations. Namely, a network works most efficiently when the weight parameters are near zero. There are no sign preferences for the weights. Moreover, 
one can distinguish several classes of weights in the network. It is the result of the internal symmetries of a network model. For instance, the interchange of two units in the same hidden layer does not change the functional form of the network map. The weights that belong to the same class should have the same scaling property \cite{Bishop_book}. According to that, we assume the following Gaussian prior density for the weights:
\begin{equation}
\label{Eq:posterior_appendix}
p(\mathbf{w}\mid \boldsymbol \alpha,\mathcal{N})
= \frac{1}{Z_W} \exp\left(-\sum_{i=1}^C \alpha_i E_w^i \right) \equiv 
\frac{1}{Z_W} \exp\left(- E_w \right),
\end{equation}
and
\begin{eqnarray}
E_w^i &=& \frac{1}{2} \displaystyle\sum_{k = s_{i-1}}^{ s_{i}} w_{k}^2, 
\quad s_i =  \sum_{l=1}^{i} W_l, \; s_0=1
\\ 
Z_W(\boldsymbol \alpha) &=& \prod_{i=1}^C \left( \frac{2\pi}{\alpha_i}\right)^\frac{W_i}{2},
\end{eqnarray}
In the above, $C$ denotes the number of weight classes; $w_j$ refers to the $j$-th weight, $W_i$ is the number of weights in $i$-th class; $\alpha_i$ is a decay width for $i$-th class of weights. The vector $\boldsymbol \alpha = (\alpha_1,\dots,\alpha_C)$ contains  decay weights.   

Note that the $\alpha_i$'s play two roles. They control the underfitting/overfitting. If $\alpha$ is too small, the network tends to overfit, whereas when $\alpha$ is too large, the network does not adapt to the data. The $1/\sqrt{\alpha_i}$'s define the width of the Gaussian prior for weights.  

Now, we construct the likelihood to include all contributions from the PINN loss terms. Assume that we have the constraints coming from the differential equations and one determined by the boundary condition, then we have two corresponding hyperparameters $\beta_{q}$, $\beta_b$, and the likelihood reads
\begin{equation}
p(\mathcal{D} \mid \mathbf{w} , \boldsymbol \beta ) = \frac{1}{Z_D(\boldsymbol \beta)} \exp\left( - \beta_{q} E_D^{q} - \beta_{b} E_D^{b} \right),
\end{equation}
where the normalization factor reads
\begin{equation}    
Z_D(\boldsymbol \beta) = Z_D(\beta_{q}) Z_D(\beta_b)
\end{equation}
and 
\begin{equation}
Z_D(\beta_{q (b)})  =  \left( \frac{2\pi}{\beta_{q(b)}} \right)^{\displaystyle \frac{N_{q(b)}}{2}}.
\end{equation}
Having the posterior and the likelihood, we compute the posterior for the weights
\begin{eqnarray}
p(\mathbf{w} \mid \mathcal{D},\boldsymbol \alpha, \boldsymbol \beta) & = & \frac{p(\mathcal{D}\mid \mathbf{w},\boldsymbol\beta) p(\mathbf{w},\boldsymbol \alpha)}{ \displaystyle \int d\mathbf{w} \, p(\mathcal{D}\mid \mathbf{w},\boldsymbol \beta) p(\mathbf{w},\boldsymbol \alpha)}
\nonumber\\
&=& \frac{\displaystyle \exp\left( - \left\{ \sum_{i=1}^C \alpha_i E_w^i + \beta_{q} E_D^{q} + \beta_{b} E_D^{b} \right\}\right) }{\displaystyle Z_T(\boldsymbol \alpha,\boldsymbol \beta)}, \nonumber \\
\end{eqnarray}
where 
\begin{eqnarray}
\label{Eq:ET_definition_appendix}
E_T &=& \sum_{i=1}^C \alpha_i E_w^i + \beta_{q} E_D^{q} + \beta_{b} E_D^{b}
\\
Z_T(\boldsymbol \alpha,\boldsymbol \beta ) 
& = &
\int d\mathbf{w} \exp\left( -E_T(\mathbf{w}) \right).
\end{eqnarray}

Then it is assumed that the posterior has a sharp local maximum and to estimate it, one expands the $E_T$ around its minimum  
\begin{equation}
E_T(\mathbf{w})  \approx E_T(\mathbf{w}_{MP}) + \frac{1}{2} \left(\mathbf{w} - \mathbf{w}_{MP} \right)^T \mathbf{A} \left(\mathbf{w} - \mathbf{w}_{MP} \right),
\end{equation}
where $\mathbf{A}$ is defined by
\begin{equation}
\mathbf{A}  = \mathbf{H} + \mathbf{I}_{W}({\boldsymbol\alpha}),
\end{equation}
where
\begin{equation}
\mathbf{I}_{W}({\boldsymbol\alpha}) = \mathrm{diag}(\underbrace{\alpha_1,\dots,\alpha_1}_{W_1\, times},\dots, \underbrace{\alpha_C,\dots,\alpha_C}_{W_C\, times})
\end{equation}
and 
\begin{equation}
\mathbf{H}_{ij} = \beta_{q} \nabla_i \nabla_j E_D^{q} + \beta_{b} \nabla_i \nabla_j E_D^{b}.
\end{equation}

Within the above approximation, one gets:
\begin{widetext}
\begin{equation}
p(\mathbf{w}\mid \mathcal{D},\boldsymbol\alpha, \boldsymbol \beta) 
= \frac{1}{Z_{E_T}^*} 
\exp\left( - E_T(\mathbf{w}_{MP}) -\frac{1}{2} \left(\mathbf{w} - \mathbf{w}_{MP} \right)^T \mathbf{A} \left(\mathbf{w} - \mathbf{w}_{MP} \right)  \right),
\end{equation}
\end{widetext}
where $Z_{E_T} \to Z_{E_T}^*$,
\begin{equation}
Z_{E_T}^*  = \exp\left( - E_T(\mathbf{w}_{MP}) \right) \sqrt{\frac{(2\pi)^W}{|\mathbf{A}|}}.
\end{equation}
$|\mathbf{A}|$ is determinant of matrix $\mathbf{A}$. Eventually, we obtain the posterior
\begin{widetext}
\begin{equation}
p(\mathbf{w}\mid \mathcal{D},\boldsymbol\alpha,\boldsymbol\beta) 
= \sqrt{\frac{|\mathbf{A}|}{(2\pi)^W}} 
\exp\left(  -\frac{1}{2} \left(\mathbf{w} - \mathbf{w}_{MP} \right)^T \mathbf{A} \left(\mathbf{w} - \mathbf{w}_{MP} \right)  \right).
\end{equation}
\end{widetext}

Note that to find the $\mathbf{w}_{MP}$, we shall find the minimum of the loss (\ref{Eq:ET_definition_appendix}).

\subsection{Second step}

Given optimal configuration of $\mathbf{w}_{MP}$ should have corresponding optimal hyperparameters $\boldsymbol{\alpha}_{MP}$, $\boldsymbol{\beta}_{MP}$ which sit in local maximum of  the posterior $p(\boldsymbol \alpha, \boldsymbol \beta \mid \mathcal{N}, \mathcal{D} )$. We have limited prior knowledge about the hyperparameters, so non-informative priors are discussed \cite{berger1985statistical}. Usually, all (positive) hyperparameter values are assumed to be valid. As a result, the likelihood density $p(\mathcal{D} \mid \boldsymbol \alpha, \boldsymbol \beta, \mathcal{N} )$, for given optimal value of weights,  should have the peak in the same position as the posterior $p(\boldsymbol \alpha, \boldsymbol \beta \mid \mathcal{N}, \mathcal{D} )$.

Below we shall obtain hyperparameters into so-called evidence approximation~\cite{Gull1988,MacKay1992.4.3.415}. Taking into account all the above, as well as the results of the previous step, one estimates the integral
\begin{eqnarray}
p(\mathcal{D}\mid \boldsymbol \alpha, \boldsymbol\beta,\mathcal{N})
&=&
\int d\mathbf{w} p(\mathcal{D}\mid \mathbf{w}, \boldsymbol\beta)
p(\mathbf{w} \mid \boldsymbol\alpha,\mathcal{N}) \nonumber \\
&=& \frac{Z_{E_T}^*}{Z_{D}(\boldsymbol \beta) 
	Z_{W}(\boldsymbol \alpha )}, 
\end{eqnarray}
where it was assumed that prior depends only on $\boldsymbol\alpha$ whereas the likelihood on $\boldsymbol \beta$ only. Then one gets
\begin{eqnarray}
\ln  p(\mathcal{D}\mid \boldsymbol\alpha,\boldsymbol\beta,\mathcal{N}) &\approx & - \sum_{i=1}^C \alpha_i E_w^i(\mathbf{w}_{MP}) 
\nonumber \\
&      & \!\!\!\!\!\!\!\! - \beta_{q} E_D^{q}(\mathbf{w}_{MP}) - \beta_{b} E_D^{b}(\mathbf{w}_{MP})
\nonumber \\
&      & \!\!\!\!\!\!\!\!
- \frac{1}{2}\ln |\mathbf{A}| - \sum_{i=1}^C \frac{W_i}{2}\ln\left( \frac{2\pi}{\alpha_i}\right)  \nonumber \\
&  & \!\!\!\!\!\!\!\! -  \frac{N_{q}}{2} \ln \left( \frac{2\pi}{\beta_{q}} \right)
- \frac{N_{b}}{2} \ln \left( \frac{2\pi}{\beta_{b}} \right). 
\end{eqnarray}
We search for the maximum of $p(\mathcal{D}\mid \boldsymbol \alpha,\boldsymbol \beta,\mathcal{N})$. In the original MacKay's framework, the hyperparameters are iterated according to the necessary condition for the maximum, namely, from the equations:
\begin{eqnarray}
\frac{\partial}{ \partial \boldsymbol\alpha} p(\mathcal{D}\mid \boldsymbol \alpha, \boldsymbol\beta,\mathcal{N}) =0, 
\\
\frac{\partial}{ \partial \boldsymbol\beta} p(\mathcal{D}\mid \boldsymbol \alpha, \boldsymbol\beta,\mathcal{N}) =0.
\end{eqnarray}
However, the above equations were solved approximately, and, in particular, the second equation led to instabilities in the iteration scheme. Hence, in our approach, we propose to consider additional loss given by $E_{hyp} = -\ln  p(\mathcal{D}\mid \boldsymbol \alpha,\boldsymbol \beta)$ and optimize it with respect to hyperparameters. The loss, after neglecting constant terms, has the form
\begin{eqnarray}
E_{hyp}  &=&   \sum_{i=1}^C \alpha_i E_w^i(\mathbf{w}_{MP}) \nonumber \\
& & + \beta_{q} E_D^{q}(\mathbf{w}_{MP}) + \beta_{b} E_D^{b}(\mathbf{w}_{MP})
+ \frac{1}{2}\ln |\mathbf{A}|  
\nonumber \\
&     & 
-\sum_{i=1}^C \frac{W_i}{2}\ln\alpha_i
-  \frac{N_{q}}{2} \ln \beta_{q}
- \frac{N_{b}}{2} \ln \beta_{b} .
\label{Eq:HyperLoss_II_Appendix}
\end{eqnarray}

\subsection{Third step}

To compute the evidence, one has to integrate out the hyperparameters from $p(\mathcal{D}\mid \boldsymbol \alpha, \boldsymbol\beta,\mathcal{N})$.  using the similar arguments as for the prior for the weights, namely, the posterior for the hyperparameters has a sharp peak in their parameters. As it is argued in \cite{Bishop_book,berger1985statistical}, the scaling parameters are always positive. Therefore, instead of considering $p(\mathcal{D}\mid \boldsymbol \alpha, \boldsymbol\beta,\mathcal{N}) \boldsymbol d \alpha \boldsymbol d\beta$, one discusses $p(\mathcal{D}\mid \ln\boldsymbol \alpha, d\ln\boldsymbol\beta,\mathcal{N}) d\boldsymbol \ln \alpha \boldsymbol \ln \beta$. Then it is assumed that the prior for $\ln$ of hyperparameters is flat. As a result, the log of evidence reads
\begin{eqnarray}
\ln p(\mathcal{D}\mid \mathcal{N}) 
&=&     
\ln  p(\mathcal{D}\mid \boldsymbol\alpha,\boldsymbol\beta) + \ln \sigma_{\ln\alpha} \nonumber \\
& & +  \ln \sigma_{\ln\beta_{q}} + \ln \sigma_{\ln\beta_b}  + comb, 
\label{Eq:Evidence_general_Ia_Appendix}
\end{eqnarray}
where
\begin{widetext}
\begin{eqnarray}
\ln  p(\mathcal{D}\mid \boldsymbol\alpha,\boldsymbol\beta) &=& 
- \sum_{i=1}^C \alpha_i E_w^i(\mathbf{w}_{MP}) 
- \beta_{Eq} E_D^{q}(\mathbf{w}_{MP}) - \beta_{b} E_D^{b}(\mathbf{w}_{MP})
- \frac{1}{2}\ln |\mathbf{A}|   +\sum_{i=1}^C \frac{W_i}{2}\ln\alpha_i
+  \frac{N_{q}}{2} \ln \beta_{q}
+ \frac{N_{b}}{2} \ln \beta_{b} 
\nonumber
\\
\label{Eq:final_appendix}
\end{eqnarray}
\end{widetext}
and
\begin{eqnarray}
\ln {\sigma_{\ln\alpha_i}} &=& -\frac{1}{2} \ln \left[ -
\alpha^2_i \frac{\partial^2 }{\partial \alpha_i^2} \ln p(\mathcal{D}\mid \boldsymbol\alpha, \boldsymbol\beta ) \right],
\\
\ln {\sigma_{\ln\beta_i}} &=& -\frac{1}{2} \ln \left[ -
\beta^2_i \frac{\partial^2 }{\partial \beta_i^2} \ln p(\mathcal{D}\mid \boldsymbol\alpha, \boldsymbol\beta ) \right].
\end{eqnarray}

\section*{Acknowledgments}

K.M.G was supported by the program ”Excellence initiative—research university” (2020-2026 University of Wroclaw).

%

%

\end{document}